%% file: main.tex
\documentclass[sigconf, nonacm, screen]{acmart}

\newtheorem{myExa}{Example}

\usepackage[ruled,vlined]{algorithm2e}
\usepackage{algpseudocode}
\usepackage{multirow}
\usepackage{stackengine}
\usepackage{makecell}
\usepackage{tcolorbox}
\usepackage{colortbl}
\usepackage{diagbox}
\usepackage{bbding}

\newcommand{\fts}{\textsf{FTS}}
\newcommand{\svs}{\textsf{SVS}}
\newcommand{\dvs}{\textsf{DVS}}
\newcommand{\ts}{\textsf{TenS}}

\newcommand{\squishlist}{
	\begin{list}{$\bullet$}
		{ \setlength{\itemsep}{1pt}
			\setlength{\parsep}{1pt}
			\setlength{\topsep}{2.5pt}
			\setlength{\partopsep}{0.5pt}
			\setlength{\leftmargin}{1em}
			\setlength{\labelwidth}{1em}
			\setlength{\labelsep}{0.6em}
		}
	}
	\newcommand{\squishend}{
	\end{list}
}

\newcommand{\rating}[1]{%
  \ifcase#1
    \FiveStarOpen\FiveStarOpen\FiveStarOpen\FiveStarOpen\FiveStarOpen 
  \or
    \FiveStar\FiveStarOpen\FiveStarOpen\FiveStarOpen\FiveStarOpen 
  \or
    \FiveStar\FiveStar\FiveStarOpen\FiveStarOpen\FiveStarOpen 
  \or
    \FiveStar\FiveStar\FiveStar\FiveStarOpen\FiveStarOpen 
  \or
    \FiveStar\FiveStar\FiveStar\FiveStar\FiveStarOpen 
  \or
    \FiveStar\FiveStar\FiveStar\FiveStar\FiveStar 
  \fi
}

\usepackage{tikz}

\makeatletter
  \newcommand\figcaption{\def\@captype{figure}\caption}
  \newcommand\tabcaption{\def\@captype{table}\caption}
\makeatother

\makeatletter
\patchcmd{\@algocf@start}
  {-1.5em}
  {0pt}
  {}{}
\makeatother

\newcommand\vldbavailabilityurl{https://github.com/whenever5225/infinity}

\newcommand\vldbpagestyle{plain}

\begin{document}

\title{Balancing the Blend: An Experimental Analysis of \\Trade-offs in Hybrid Search}

\settopmatter{authorsperrow=4}

\author{Mengzhao Wang}
\affiliation{%
  \institution{Zhejiang University}
}
\email{wmzssy@zju.edu.cn}

\author{Boyu Tan}
\affiliation{%
  \institution{Zhejiang University}
}
\email{jacktby@gmail.com}

\author{Yunjun Gao}
\affiliation{%
  \institution{Zhejiang University}
}
\email{gaoyj@zju.edu.cn}

\author{Hai Jin}
\affiliation{%
  \institution{Infiniflow}
}
\email{hai.jin@infiniflow.ai}

\author{Yingfeng Zhang}
\affiliation{%
  \institution{Infiniflow}
}
\email{yingfeng.zhang@infiniflow.ai}

\author{Xiangyu Ke}
\affiliation{%
  \institution{Zhejiang University}
}
\email{xiangyu.ke@zju.edu.cn}

\author{Xiaoliang Xu}
\affiliation{%
  \institution{Hangzhou~\mbox{Dianzi}~\mbox{University}}
}
\email{xxl@hdu.edu.cn}

\author{Yifan Zhu}
\affiliation{%
  \institution{Zhejiang University}
}
\email{xtf_z@zju.edu.cn}

\begin{abstract}
Hybrid search, the integration of lexical and semantic retrieval, has become a cornerstone of modern information retrieval systems, driven by demanding applications like Retrieval-Augmented Generation (RAG). The architectural design space for these systems is vast and complex, yet a systematic understanding of the trade-offs among their core components—retrieval paradigms, combination schemes, and re-ranking methods—is lacking. To address this, and informed by our experience building the \texttt{Infinity} open-source database, we present the first experimental analysis of advanced hybrid search architectures. Our framework integrates four retrieval paradigms—Full-Text Search (\fts), Sparse Vector Search (\svs), Dense Vector Search (\dvs), and Tensor Search (\ts)—and evaluates their combinations and re-ranking strategies across 11 real-world datasets. Our results reveal three key findings: (1) A ``weakest link'' phenomenon, where a weak path can substantially degrade overall accuracy, highlighting the need for path-wise quality assessment before fusion. (2) A data-driven map of performance trade-offs, demonstrating that optimal configurations depend heavily on resource constraints and data characteristics, precluding a one-size-fits-all solution. (3) The identification of Tensor-based Re-ranking Fusion (TRF) as a high-efficacy alternative to mainstream fusion methods, offering the semantic power of tensor search at a fraction of the computational and memory cost. Our findings offer concrete guidelines for designing adaptive, scalable hybrid search systems and identify key directions for future research.

\end{abstract}

\maketitle

\pagestyle{\vldbpagestyle}

\ifdefempty{\vldbavailabilityurl}{}{
\begingroup\small\noindent\raggedright\textbf{PVLDB Artifact Availability:}\\
The source code, data, and/or other artifacts have been made available at \url{\vldbavailabilityurl}.
\endgroup
}

\input{sections/1_introduction}
\input{sections/2_background}
\input{sections/3_framework}
\input{sections/4_eval_setup}
\input{sections/5_results}
\input{sections/6_discussion}
\input{sections/7_conclusion}

\begin{acks}
The evaluation framework presented in this paper was significantly informed by the architectural principles and practical experience gained from the development of the \texttt{Infinity} open-source database\textsuperscript{\ref{repolink}}. We are deeply grateful to our colleagues on the Infinity team for their foundational work and insightful discussions. Specifically, we would like to extend our sincere thanks to Zhichang Yu, Yushi Shen, Zhiqiang Yang, Ling Qin, and Yi Xiao for their invaluable contributions to the \texttt{Infinity} project.
\end{acks}


\bibliographystyle{ACM-Reference-Format}
\bibliography{myref}

\end{document}

%% file: sections/1_introduction.tex
\section{Introduction}
\label{sec: intro}
Modern information retrieval tasks increasingly require database systems to move beyond single-paradigm search \cite{SawarkarMS24,abs-2410-20381}. Hybrid search architectures, blending multiple retrieval techniques, have emerged as a powerful solution, driven by applications like Retrieval-Augmented Generation (RAG) \cite{Dong24,Andromeda,AquaPipe}. However, designing these systems for optimal performance is challenging, as the trade-offs between retrieval quality, efficiency, and cost are complex \cite{JiangZWHA24,ZhaoZL24,WangWKGXC24}. This challenge is particularly acute for advanced architectures combining three or more paradigms, as their synergistic benefits and interference effects remain unstudied. This leaves system designers without a clear guide, raising an important question for the database community: \textbf{how can we systematically navigate this complex design space of hybrid search to make evidence-based architectural decisions?}

\begin{figure}
  \setlength{\abovecaptionskip}{-0.06cm}
  \setlength{\belowcaptionskip}{-0.34cm}
  \centering
  \footnotesize
  \hspace{0.3cm}
  \stackunder[0.5pt]{\includegraphics[scale=0.156]{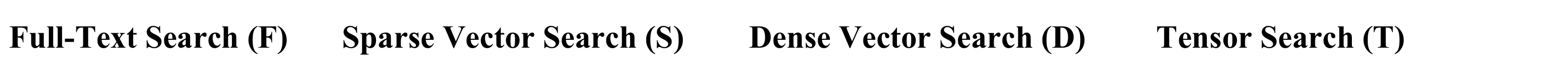}}{}
  \newline
  \stackunder[0.5pt]{\includegraphics[scale=0.189]{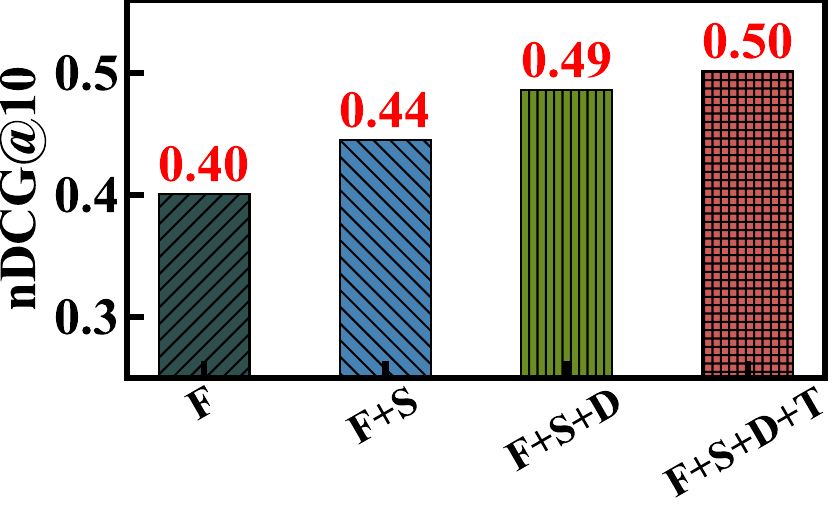}}{(a) Accuracy}
  \stackunder[0.5pt]{\includegraphics[scale=0.189]{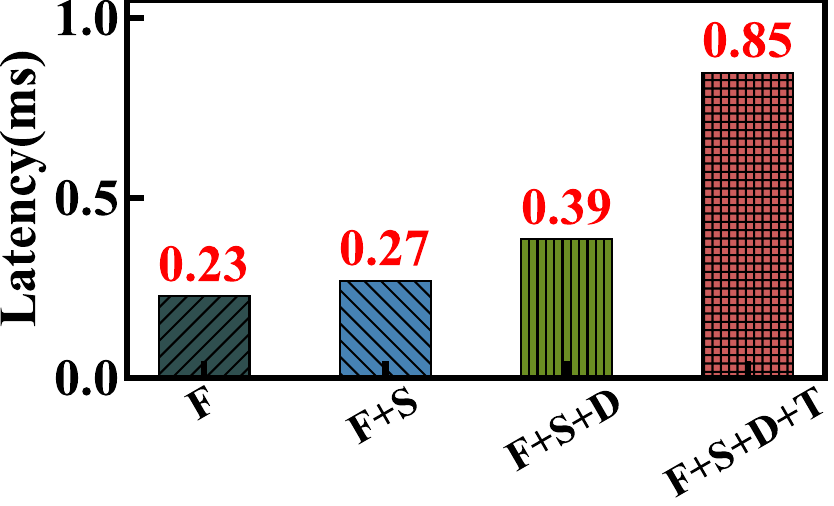}}{(b) Efficiency}
  \stackunder[0.5pt]{\includegraphics[scale=0.189]{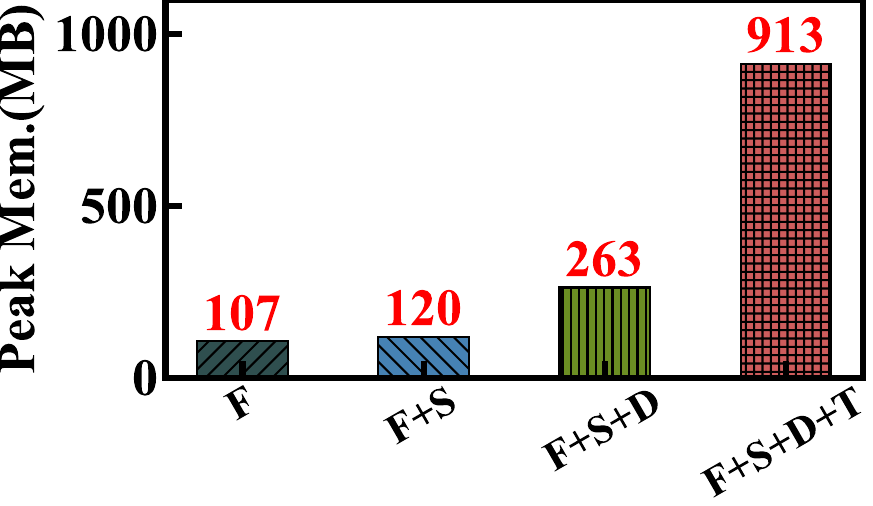}}{(c) Overhead}
  \newline
  \caption{The multi-faceted performance of hybrid search, with all metrics evaluated on the CQAD(en) dataset.}
  \label{fig: intro perf}
  \vspace{-0.2cm}
\end{figure}

The study of retrieval has a long history \cite{TurtleF95}, stemming from decades of work in both the Database (DB) \cite{ChakrabartiCG11,Starling,WangLHCPS17,WidmoserKA24} and Information Retrieval (IR) \cite{GrandMFL20,BroderCHSZ03,0003WZC0RRR24,Yang0L17} communities.
This research has produced two complementary approaches: lexical and semantic search. Lexical methods, such as Full-Text Search (\fts) \cite{WangLPS17,DingS11} and Sparse Vector Search (\svs) \cite{BruchNIL24a,FormalPC21}, excel at exact keyword matching and offer interpretability, but often fail to capture contextual meaning \cite{GyselRK18}. In contrast, semantic approaches like Dense Vector Search (\dvs) \cite{WangXY021,LiZSWLZL20} and Tensor Search (\ts) \cite{KhattabZ20,SanthanamKPZ22} employ neural models to understand context and nuance \cite{WangMW22,milvus}, though they may lack precision for keyword-specific queries. Hybrid search strategically combines these paradigms to synthesize their complementary strengths \cite{ChenZLBN22,MaKHM20,abs-1903-08690}. While this blended approach can significantly improve retrieval accuracy, it introduces a fundamental tension between effectiveness and system cost. As our preliminary results in Figure \ref{fig: intro perf} demonstrate, each additional retrieval path, while boosting accuracy, can dramatically increase query latency and memory consumption. This creates a complex optimization problem, making it crucial for system designers to carefully balance these competing factors.

This challenge, however, is exacerbated by several critical research gaps.
First, existing hybrid search studies \cite{LuanETC21,abs-2410-20381,abs-2402-03216} are often narrow in scope, typically focusing on pairwise combinations; a systematic understanding of how three or more paradigms perform and interfere in a single system is lacking.
Second, common re-ranking strategies rely on simple fusion methods like Reciprocal Rank Fusion (RRF) \cite{CormackCB09,BruchGI24}, whose effectiveness is unverified for complex scenarios where candidate lists from multiple paradigms must be consolidated.
Finally, and most critically, the community lacks a framework to systematically evaluate and compare the trade-offs of all four major paradigms (full-text \cite{WangLHCPS17}, sparse \cite{0006CHZZ024}, dense \cite{GaoL23}, and tensor \cite{NardiniRV24}), which hinders reproducible research.
Given the proven, yet distinct, value of each paradigm \cite{QiaoYL023,GaoL23,0006CHZZ024,SawarkarMS24,NardiniRV24}, such a study is essential for understanding their intrinsic trade-offs and guiding future system design.

To bridge these gaps, this paper presents the first comprehensive experimental analysis of advanced hybrid search architectures. We built a modular evaluation framework, informed by our work on the \texttt{Infinity}\footnote{\url{https://github.com/infiniflow/infinity} \label{repolink}} open-source database, supporting the systematic evaluation of arbitrary combinations of the four retrieval paradigms. Using this framework, we conduct an experimental study across 11 real-world datasets to create a data-driven map of the hybrid search performance landscape. Our investigation is guided by three central research questions: (\textbf{RQ1}) Does integrating additional retrieval paths consistently improve search accuracy? (\textbf{RQ2}) What criteria guide the selection of appropriate combination schemes for different scenarios? (\textbf{RQ3}) How can candidates from different retrieval paths be effectively re-ranked within a database?

\textbf{Key Findings and Implications.} Our experimental analysis yields three key findings that provide practical guidelines for system builders and new directions for researchers:
\textbf{First}, we identify a ``weakest link'' phenomenon in multi-path architectures. While combining paths can improve accuracy, our results provide the first systematic evidence that a weak path can \emph{substantially} degrade overall performance. This highlights the critical need for path-wise quality assessment before fusion.
\textbf{Second}, we provide a data-driven map of performance trade-offs, demonstrating that no single hybrid architecture excels across accuracy, efficiency, and resource cost simultaneously. Rather, the ideal configuration is dependent on specific constraints (e.g., latency, memory) and data characteristics; our analysis provides quantitative evidence to guide these design choices.
\textbf{Third}, we demonstrate that Tensor-based Re-ranking Fusion (TRF) is a highly effective and practical re-ranking strategy. It consistently outperforms mainstream fusion methods like Reciprocal Rank Fusion (RRF), offering the semantic power of a full tensor search at a fraction of the computational and memory cost.

Our major contributions are as follows:

\squishlist

\item \textbf{A Systematic Survey of Retrieval Paradigms}. We present the first systematic survey comparatively analyzing four major retrieval paradigms. We analyze their historical evolution, technical underpinnings, and respective trade-offs, establishing the foundation for advanced hybrid search.

\item \textbf{A Novel Open-Source Evaluation Framework}. We design, build, and open-source a modular framework for evaluating advanced hybrid search. It is the first publicly available tool supporting the flexible combination and evaluation of all four major retrieval paradigms and multiple re-ranking strategies.

\item \textbf{A Comprehensive Evaluation and Data-Driven Analysis}. Using our framework, we conduct a rigorous evaluation of retrieval combinations and re-ranking strategies across 11 established real-world datasets. Our analysis quantifies the trade-offs in accuracy, efficiency, and operational overhead, offering concrete, evidence-based guidelines for practitioners.

\item \textbf{Identification of Key Challenges and Future Directions}. Based on our empirical findings, we provide observations to guide scenario-specific architectural choices and articulate key challenges and promising research directions for hybrid search.

\squishend

The remainder of this paper is organized as follows. We review retrieval paradigms in Section \ref{sec: background} and detail our evaluation frame- work in Section \ref{sec: framework}. We then describe the experimental setup in Section \ref{sec: eval setup} and present our comprehensive results in Section \ref{sec: exp res}. We discuss our findings, their implications, and future work in Section \ref{sec: discussion}. Section \ref{sec: conclusion} concludes the paper.

%% file: sections/2_background.tex
\section{Background}
\label{sec: background}
This section reviews the technical underpinnings of the four major paradigms that constitute modern hybrid search.

\subsection{Lexical Search Paradigms}
Lexical search paradigms operate on the textual representation of documents and queries, determining relevance primarily through keyword frequency and statistical signals. While valued for their efficiency, interpretability, and precision, their effectiveness is constrained by the ``vocabulary mismatch'' problem \cite{GangulyRMJ15}. Treating text as an unordered collection of words, they often struggle to capture contextual nuance. This section reviews the two most prominent lexical paradigms---traditional Full-Text Search ({\fts}) and modern learned Sparse Vector Search ({\svs})---whose distinct characteristics are illustrated in Figure \ref{fig:retrieval method example} (left).

\subsubsection{Full-Text Search (\fts)}
\begin{figure}
  \centering
  \setlength{\abovecaptionskip}{0.1cm}
  \setlength{\belowcaptionskip}{-0.3cm}
  \includegraphics[width=\linewidth]{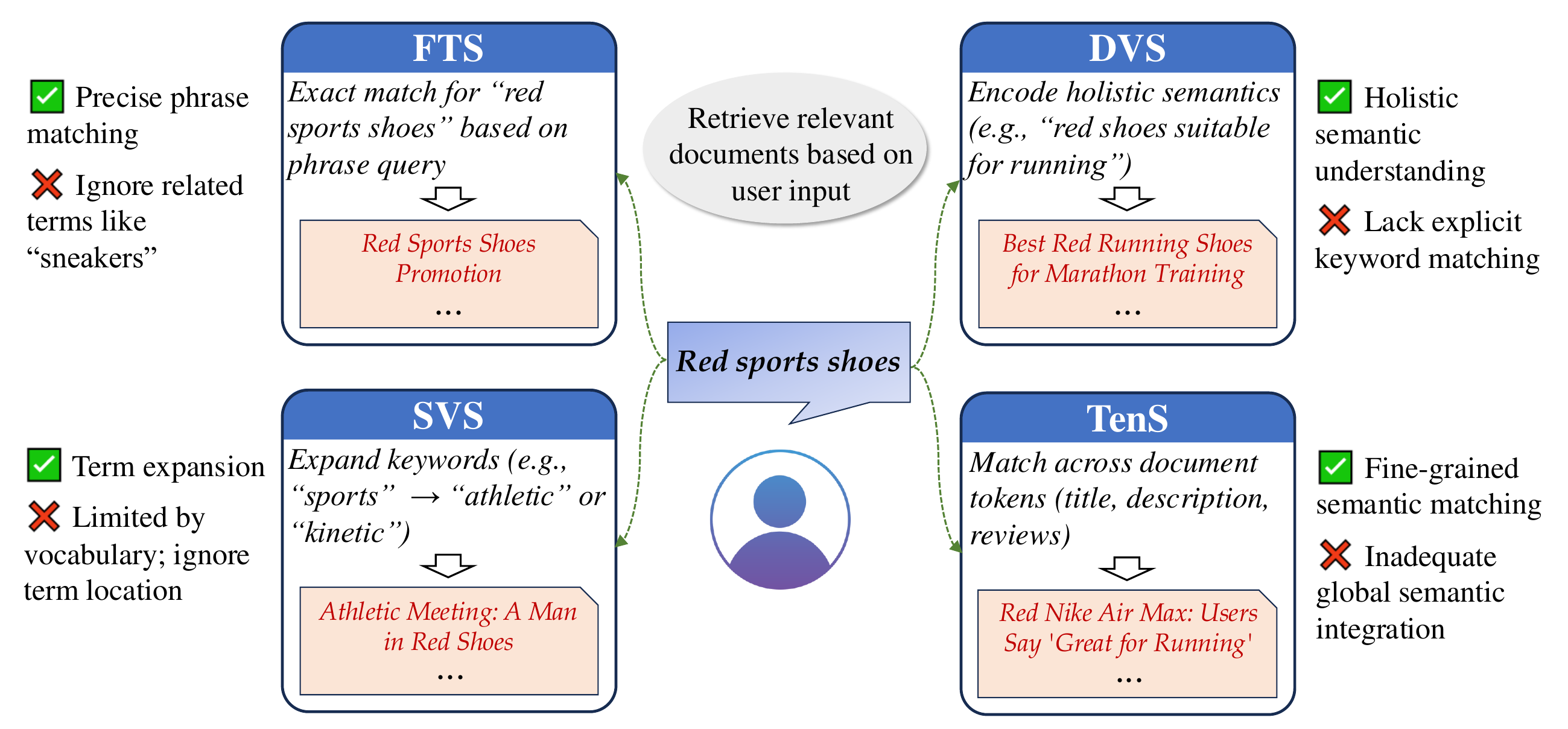}
  \caption{A toy example of different retrieval paths.}
  \label{fig:retrieval method example}
\end{figure}

{\fts} is a foundational lexical retrieval technology that has evolved through decades of optimization. Its development progressed from early Boolean models, which offered efficient filtering but lacked relevance scoring \cite{Radecki88,PohlMZ12,HristidisHI10}, to statistical weighting schemes like TF-IDF that introduced the concept of term importance \cite{AmatiR02,PaulsenGD23,Paik13}. This formed the basis for modern probabilistic models, with the Okapi BM25 (Best Matching 25) algorithm now the de-facto standard for relevance ranking in production systems \cite{TrotmanPB14,RobertsonWJHG94,HadjieleftheriouKS09,TrotmanK11}, such as Elasticsearch~\cite{elasticsearch} and OpenSearch~\cite{opensearch}. BM25 enhances earlier frameworks by incorporating two key heuristics: non-linear term frequency saturation, which prevents overly frequent terms from dominating the relevance score, and document length normalization, which adjusts scores for varying document sizes. Given a query $Q$ and a document $D$, the BM25 score is calculated as:
\begin{equation}
\label{equ:bm25}
sim(Q,D)=\sum_{i=1}^{n}IDF(q_i) \cdot \frac{f(q_i, D) \cdot (k_1 +1 )}{f(q_i, D)+k_1 \cdot (1 - b + b \cdot \frac{len(D)}{avgdl})} ,
\end{equation}
where $f(q_i, D)$ is the term frequency of query term $q_i$ in document $D$, $len(D)$ is the document length, and $avgdl$ is the average document length. The parameters $k_1$ and $b$ control term frequency saturation and length normalization, respectively. A higher score indicates greater relevance.
This reliance on exact term statistics defines {\fts} behavior. As illustrated in Figure \ref{fig:retrieval method example} (top left), for a query like ``\textsf{red sports shoes}'', {\fts} excels at precise phrase matching. This precision is a key strength, but the reliance on exact terms also reveals its primary limitation: a lack of semantic understanding, causing it to fail on synonyms like ``\textsf{sneakers}'' for ``\textsf{shoes}''.

For large-scale retrieval, {\fts} engine efficiency and effectiveness rely heavily on advanced dynamic pruning \cite{GrandMFL20,TurtleF95,MalliaOPTV17,TonellottoMO18,CraneCLMT17} and flexible query capabilities \cite{Sack2008,renouf2007webcorp}. Pruning algorithms such as WAND (Weak AND) \cite{BroderCHSZ03} and its modern successor, Block-Max WAND \cite{DingS11}, enable early termination by calculating real-time score bounds, effectively filtering low-value candidates without full scoring. Beyond ranking, {\fts} supports rule-based query operations (e.g., phrase searches, wildcards, term weighting) that enable highly controllable retrieval. These capabilities, combined with optimizations like inverted index compression \cite{MalliaSS19,PibiriV21}, form the technical stack enabling modern {\fts} engines to handle massive corpora.

\subsubsection{Sparse Vector Search (\svs)}
{\svs} is a modern lexical search approach that enhances traditional term-based methods with learned term weights from deep neural models \cite{abs-2106-14807,abs-2402-03216}. Unlike {\fts}, which relies on raw corpus statistics, {\svs} employs learning models like SPLADE \cite{FormalPC21,SPLADE-v2,SPLADE-v3} to transform text into high-dimensional sparse vectors. In these vectors, each dimension corresponds to a term in an expanded vocabulary, and non-zero values represent the learned contextual importance of that term. Relevance between a query $Q$ and a document $D$ is computed as the inner product (IP) of their sparse vectors, $\boldsymbol{q}$ and $\boldsymbol{d}$:
\begin{equation}
\label{eq: inner product}
sim(Q, D) = \boldsymbol{q}^{\top} \boldsymbol{d} = \sum_{i=1}^{n} q_i d_i,
\end{equation}
A higher IP indicates greater relevance. This learning-based approach grants {\svs} a key advantage over {\fts}: term expansion. As shown in Figure \ref{fig:retrieval method example} (bottom left), {\svs} can map ``\textsf{sports}'' to related terms like ``\textsf{athletic}'', mitigating the vocabulary mismatch problem to a degree. However, {\svs} remains a lexical method. It is constrained by its vocabulary and disregards term position, which limits its ability to capture complex phrasal or high-level semantics.

The high dimensionality of sparse vectors makes efficient search challenging, a problem addressed using techniques analogous to those in {\fts}. {\svs} systems build inverted indexes on the non-zero vector dimensions, applying dynamic pruning algorithms to accelerate query processing \cite{MackenzieMMP22,MackenziePM22,abs-2501-11628,abs-2502-14822,BruchNRV24,BruchNRV24-cikm24,NguyenMY23}. Advanced methods such as Block-Max Pruning (BMP) \cite{BMP} precompute maximum term scores for document blocks, allowing the system to prune large portions of the search space without full evaluation. The efficacy of this method is demonstrated by its adoption in industrial systems like Pinecone~\cite{pinecone}. By combining learned relevance with fast pruning, {\svs} offers a powerful paradigm that bridges the gap between traditional keyword statistics and deeper semantic understanding.

\subsubsection{Comparison and Synthesis: {\fts} vs. {\svs}}
While both {\fts} and {\svs} are lexical paradigms that rely on inverted indexes and pruning, their foundational differences in relevance modeling create distinct trade-offs. The core distinction is the origin of term weights: {\fts} derives relevance from unsupervised corpus statistics (e.g., BM25), making it robust and training-free. In contrast, {\svs} uses pre-trained neural models to learn contextual term importance, capturing nuances statistical methods miss. This leads to a key practical trade-off: {\fts} provides highly controllable, rule-based operations (e.g., precise phrase and wildcard searches), while {\svs} excels at learned term expansion, mitigating vocabulary mismatch at the cost of direct control. From a systems perspective, this also introduces a significant difference in operational overhead: {\fts} indexes raw text directly, whereas {\svs} requires a computationally intensive inference step for vector generation before indexing. Ultimately, the choice balances the statistical robustness and controllability of {\fts} against the contextual matching of {\svs}, highlighting their complementary nature.

\subsection{Semantic Search Paradigms}
In contrast to lexical methods, semantic search paradigms employ deep neural networks to transform text into rich numerical representations that capture context. Their primary strength is the ability to understand synonyms, related concepts, and high-level semantic intent. This semantic depth, however, can come at the cost of keyword precision. Furthermore, reliance on pre-trained models introduces significant computational overhead and challenges in domain adaptation. This section examines the two dominant semantic paradigms: {\dvs}, which models text with a single global vector, and {\ts}, which uses a multi-vector representation. We illustrate their distinct characteristics in Figure \ref{fig:retrieval method example} (right).

\subsubsection{Dense Vector Search (\dvs)}
{\dvs} represents a text's entire semantic meaning (e.g., a sentence or document) as a single, fixed-dimensional dense vector, or embedding. Embeddings are typically generated using bi-encoder architectures \cite{MaKYHM21,DPR,thakur2beir}, where models like Sentence-BERT \cite{ReimersG19} aggregate token-level representations into a holistic summary, such as by using a special token's output (e.g., [\texttt{CLS}] \cite{abs-2402-03216,ConneauKGCWGGOZ20}) or mean-pooling all token hidden states \cite{PuccettiMD21}. Semantic similarity between a query $Q$ and a document $D$ is then computed as the inner product of their dense vectors; a higher score indicates greater similarity. This global representation enables a holistic semantic understanding, as illustrated in Figure \ref{fig:retrieval method example} (top right). For the query ``\textsf{red sports shoes}'', {\dvs} can retrieve documents about ``\textsf{running shoes}'' by capturing the query's high-level intent, even without the exact keywords. However, this aggregation process obscures fine-grained details, leading to reduced precision for queries dependent on exact terms.

Exact search over millions or billions of dense vectors is computationally prohibitive. Consequently, {\dvs} systems rely on Approximate Nearest Neighbor Search (ANNS) algorithms \cite{tauMG,WangWCWPW24,CaiSCZ24,ZuoQZLD24,PanWL24,AziziEP25} to achieve low-latency, high-throughput retrieval. ANNS algorithms include tree-based \cite{KDTree,VPTree}, hashing-based \cite{GaoJLO14,HuangFZFN15,LiuCHLS14,LSH}, quantization-based \cite{AndreKS15,AguerrebereBHTW23,GaoL24,PQ}, and graph-based approaches~\cite{DiskANN,NSG,WangXY021}. Graph-based algorithms, particularly HNSW (Hierarchical Navigable Small World graph) \cite{HNSW}, have recently emerged as the state-of-the-art, offering a superior balance of accuracy and efficiency. These techniques form the core engine of most modern vector databases~\cite{ChenJZPWWHSWW24,milvus,PASE}, including SingleStore~\cite{ChenJZPWWHSWW24}, Chroma~\cite{chroma}, and Qdrant~\cite{qdrant}.

\subsubsection{Comparison and Synthesis: {\svs} vs. {\dvs}}
While both {\svs} and {\dvs} employ vector representations, they embody fundamentally different philosophies of semantic modeling. The primary distinction is representation granularity. {\svs} focuses on term-level semantic expansion, creating high-dimensional vectors where each dimension corresponds to a vocabulary term with a learned weight. This retains a strong lexical grounding, evolving traditional ``bag-of-words'' models. In contrast, {\dvs} pursues holistic semantic summarization, compressing a text's entire meaning into a single dense vector where individual dimensions lack explicit meaning, allowing it to capture high-level intent. This representational difference dictates their underlying architectures: {\svs} typically utilizes the classic inverted index and dynamic pruning stack from {\fts}, while {\dvs} relies on an entirely different class of ANNS algorithms. These differences make them highly complementary; {\svs} offers a powerful, learned extension of lexical search, while {\dvs} provides a purely semantic pathway, explaining their frequent pairing in modern hybrid search systems. For instance, vector databases like Milvus~\cite{hybrid-search-milvus} and Weaviate~\cite{hybrid-search-weaviate} integrate {\svs} alongside their native {\dvs} capabilities specifically to improve retrieval accuracy on queries that require precise keyword matching.

\subsubsection{Tensor Search (\ts)}
{\ts} represents a shift from single-vector representations, instead modeling a text as a set of contextualized embeddings for each of its tokens. Pioneered by models like ColBERT \cite{KhattabZ20,ColBERTv2,LouisSDS25,answerai-colbert-small-v1,abs-2407-20750,abs-2408-16672}, {\ts} employs a ``late-interaction'' architecture. Rather than aggregating a text's meaning into one vector, it retains a tensor of token embeddings for both query and document, performing fine-grained similarity computations at query time. The relevance score is typically calculated using a max-similarity (MaxSim) operation, which aggregates the maximum similarity for each query token against all document tokens:
\begin{equation}
\label{equ: maxSim}
sim(Q,D) = \sum_{i=1}^N \max_{j=1}^M \boldsymbol{q}_i^{\top} \cdot \boldsymbol{d}_j,
\end{equation}
where $\boldsymbol{q}_i$ and $\boldsymbol{d}_j$ are the token embeddings of the query and document, respectively. This token-level interaction aligns each query token with its most relevant semantic counterpart in the document. As depicted in Figure \ref{fig:retrieval method example} (bottom right), {\ts} can match query tokens across different parts of a document; for the query ``\textsf{red sports shoes}'', it could match ``\textsf{red}'' in a title, a related token like ``\textsf{Nike}'' in its description, and ``\textsf{running}'' in a user review. This aggregation of best-token matches captures fine-grained semantic relationships. However, this focus on local token alignment is also its primary weakness: inadequate global semantic integration. Summing individual token scores may fail to capture the holistic context of a sentence or paragraph.

The primary drawback of {\ts} is its significant computational and memory overhead \cite{park-etal-2025-scv,ColBERTv2,GaoDC21,LiLOGLMY023}, stemming from the need to store and process an embedding for every token in the corpus. A naive implementation can result in index sizes orders of magnitude larger than for {\dvs}. Consequently, substantial research has focused on mitigating this high cost. Techniques range from representation compression (e.g., residual representations in ColBERTv2 \cite{ColBERTv2}) to multi-stage filtering pipelines like PLAID \cite{SanthanamKPZ22} and quantization-based methods like EMVB \cite{NardiniRV24} that use hardware acceleration (e.g., SIMD) for efficiency. Such techniques have enabled its use in industrial systems like Vespa~\cite{vespa} where accuracy is paramount. Despite these advancements, the high cost of {\ts} remains a significant barrier to widespread adoption, positioning it as a powerful but resource-intensive ``heavyweight'' semantic search paradigm.

\subsubsection{Comparison and Synthesis: {\dvs} vs. {\ts}}
While both {\dvs} and {\ts} are semantic paradigms that use neural embeddings, they differ fundamentally in their semantic granularity and resulting retrieval architectures. {\dvs} employs holistic summarization, compressing an entire text into a single, fixed-dimensional vector. This global representation is powerful for capturing high-level topics and intent but inherently sacrifices local details \cite{PuccettiMD21,DPR}. In contrast, {\ts} uses a fine-grained multi-vector approach, retaining a contextualized embedding for each token. This ``late-interaction'' mechanism can capture nuanced relationships but may struggle to form a coherent understanding of the text's overall global meaning \cite{KhattabZ20}.
This core difference in representation dictates their architectural trade-offs. {\dvs}, with its single-vector-per-document model, is well-suited for the mature ecosystem of ANNS algorithms, enabling efficient search over massive datasets. {\ts}, however, presents a far greater retrieval challenge due to the sheer volume of embeddings to be indexed and queried. This necessitates more complex multi-stage retrieval pipelines and specialized optimizations to remain computationally tractable. Ultimately, {\dvs} can be seen as the general-purpose semantic workhorse, while {\ts} represents a resource-intensive paradigm that trades scalability for fine-grained semantics. Their distinct strengths make them complementary.

\subsection{Hybrid Search Architectures.}
Lexical and semantic paradigms offer complementary strengths, yet neither is consistently sufficient on its own. Hybrid search synthesizes these capabilities by combining multiple retrieval paths, aiming to satisfy complex queries that require both keyword precision and semantic understanding—achieving an accuracy and robustness a single paradigm often cannot match. However, the central challenges lie in the strategic combination of retrieval paths and the effective merging of disparate candidate lists. This section examines these two core components: architectural combination schemes for path integration, and candidate re-ranking strategies for producing a unified result set.

\subsubsection{Retrieval-Path Combination Schemes}
Current hybrid search implementations \cite{BruchNIL24a,LuanETC21,YangCHQY24} are predominantly dual-path schemes, categorized by the primary system they extend. The vector-centric approach \cite{hybrid-search-weaviate,hybrid-search-pinecone,hybrid-search-milvus} integrates {\svs} into vector databases to augment native {\dvs} capabilities, addressing their weakness in handling precise keyword queries. Conversely, the text-centric approach integrates {\dvs} into established {\fts} engines like Elasticsearch \cite{elasticsearch}, retrofitting them with semantic capabilities \cite{XianTPL24,full-text-dense,abs-2311-18503}. Integration strategies for these dual-path systems vary: some create a unified index for joint pruning \cite{abs-2410-20381}, while others use separate, independent indexes managed by a multi-index engine \cite{ChenZCXZWHYLWLY24}. However, these approaches are typically limited to two paradigms. Comprehensive architectural studies and systems effectively integrating three or more paradigms—especially {\ts}—remain scarce.

\subsubsection{Candidate Re-ranking Strategies}
Merging candidate lists from multiple retrieval paths into a single result set requires a re-ranking strategy. A prevalent lightweight technique in modern systems is Reciprocal Rank Fusion (RRF) \cite{CormackCB09,RRF-industry}. RRF is a score-agnostic method that synthesizes ranked lists by prioritizing candidates appearing near the top of multiple lists; this is effective when raw scores (e.g., BM25, vector similarity) are not directly comparable \cite{ChenZLBN22}. The RRF score for a candidate $c$ is:
\begin{equation}
\label{equ:rrf score}
\text{RRF}(c) = \sum_{i=1}^{n} \frac{1}{\kappa + \text{rank}_i(c)},
\end{equation}
where $\text{rank}_i(c)$ is the rank of $c$ in the $i$-th list, $n$ is the number of lists, and $\kappa$ is a smoothing parameter.

An alternative, Weighted Sum (WS) \cite{BruchGI24,WS-industry}, linearly combines normalized scores from each path:
\begin{equation}
\label{equ:ws score}
\text{WS}(c) = \sum_{i=1}^{n} w_i \cdot \text{score}_i(c),
\end{equation}
where $w_i$ is the pre-defined weight and $\text{score}_i(c)$ is the candidate's score. Unlike RRF, WS is score-aware, suitable when scores are normalized and their relative importance can be estimated \cite{WangZZ21}.

\vspace{0.2em}
\textbf{Remarks.} These lightweight fusion methods contrast with heavyweight learning-based re-rankers, which are architecturally distinct and typically reserved for second-stage processing~\cite{turbopuffer-reranker,hs-milvus}. Prominent heavyweight models use \textit{cross-encoder} architectures~\cite{ZhangZLXDTLYXHZ24,KhattabZ20}, performing a computationally intensive \textit{joint encoding} of concatenated query and document tokens at query time. This deep interaction yields superior accuracy but is impractical for in-database deployment on large candidate sets due to high latency and hardware (e.g., GPU) requirements~\cite{ReimersG19}.
The \textit{late-interaction} architecture~\cite{abs-2408-16672,KhattabZ20,ColBERTv2} of tensor models is fundamentally different. It decouples the encoding process: document token embeddings are generated offline and indexed, while at query time, only a lightweight computation (e.g., MaxSim) is performed against the query's tokens. This design, separating expensive document encoding from online interaction, makes late-interaction a viable strategy for fine-grained re-ranking \textit{within the database}.
Our study focuses on this critical first stage of efficient in-database fusion. We evaluate both lightweight methods and the feasibility of using late-interaction models for this purpose, an approach whose trade-offs in complex hybrid systems are not yet well understood.

%% file: sections/3_framework.tex
\section{Evaluation Framework}
\label{sec: framework}
Answering our research questions requires a specialized evaluation framework that is (1) \textit{comprehensive}, supporting all four major retrieval paradigms; (2) \textit{modular}, for flexible combination; and (3) \textit{performant}, using state-of-the-art algorithms for a fair comparison. We built such a framework, informed by our development of the \texttt{Infinity}\textsuperscript{\ref{repolink}} open-source database.

\begin{figure}
  \centering
  \vspace{-0.08cm}
  \setlength{\abovecaptionskip}{0.1cm}
  \setlength{\belowcaptionskip}{-0.3cm}
  \includegraphics[width=\linewidth]{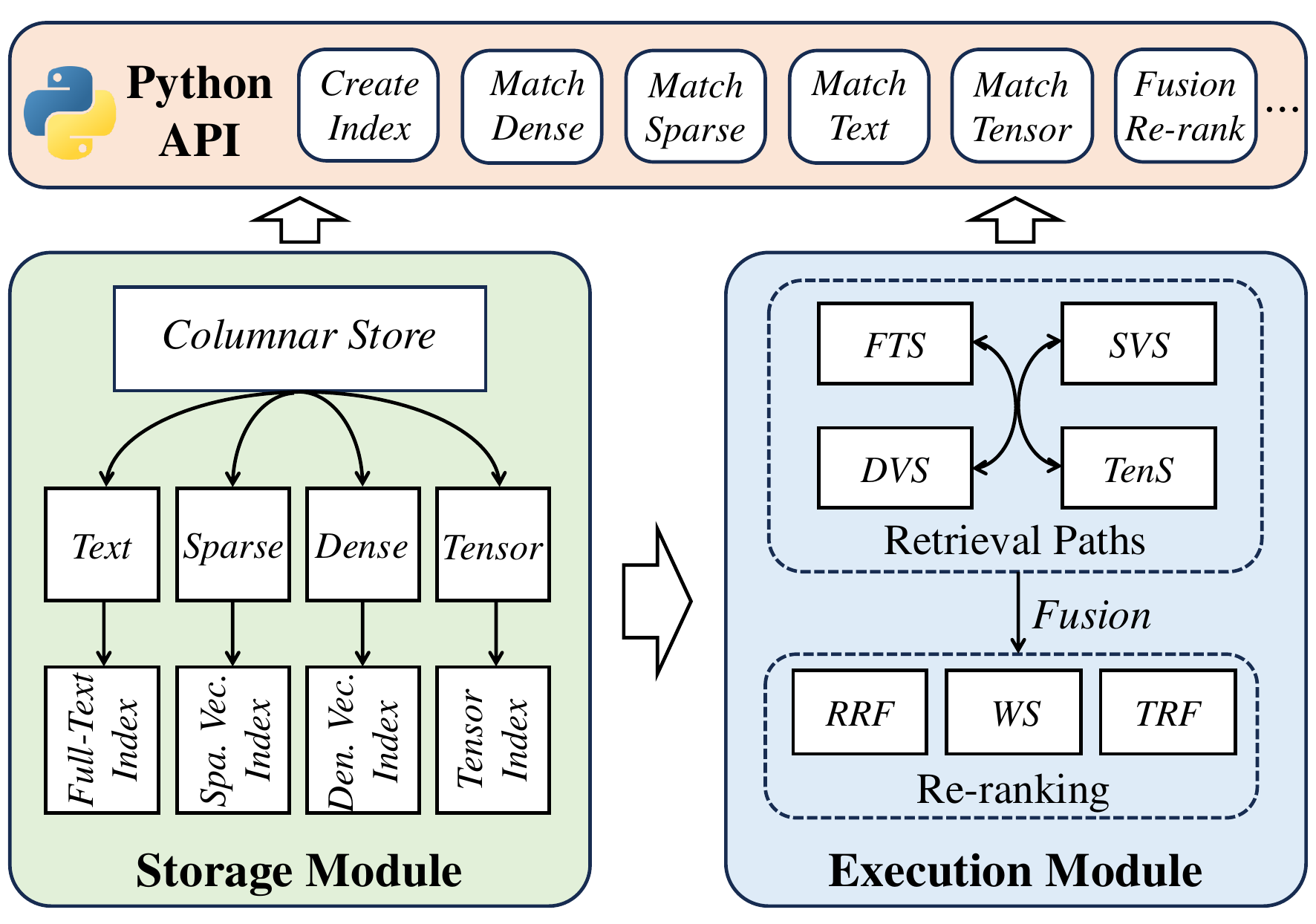}
  \caption{Overview of the evaluation framework.}
  \label{fig:framework}
\end{figure}

\subsection{Architectural Principles and Overview}
As shown in Figure \ref{fig:framework}, this framework is realized via a two-module architecture:
The \textit{Storage Module} employs a decoupled, columnar design. During ingestion, each document is partitioned into columns (Text, Sparse, Dense, Tensor), each managed by a specialized indexing fleet (e.g., HNSW for \dvs{}, optimized inverted indexes for \fts{} and \svs{}). This architectural choice is critical for modularity and performance, as each path uses its own state-of-the-art algorithm without compromise—a key limitation of unified index approaches—and ensures extensibility.
The \textit{Execution Module} orchestrates hybrid search by transforming a query into a pipelined Directed Acyclic Graph (DAG) of physical operators (see Figure~\ref{fig:dag}). Each retrieval path executes in parallel, streaming results to a final fusion operator that applies the chosen re-ranking strategy (e.g., RRF). This model is applied consistently across all combinations, ensuring a fair performance comparison. The framework is exposed via a user-friendly Python API\footnote{\url{https://infiniflow.org/docs/pysdk\_api\_reference}} for reproducibility.

\subsection{Storage Module}
\label{subsec: storage module}
All indexes are persisted to disk and accessed via a buffer manager that stages data between disk and memory, allowing the system to scale beyond main memory capacity. This disk-oriented architecture entails specific trade-offs for each paradigm:
\squishlist
\item \textbf{\fts{} and \svs{}:} These inverted-index-based paradigms are well-suited for disk storage, as their sequential access patterns within posting lists are efficiently handled by the buffer manager and OS page cache.
\item \textbf{\dvs{} (HNSW~\cite{HNSW}):} While HNSW is fastest when memory-resident, our disk-based architecture (motivated by DiskANN~\cite{DiskANN}) pages graph portions on-demand, trading higher I/O overhead for a lower memory footprint.
\item \textbf{\ts{}:} The large storage footprint of \ts{} makes a disk-based approach a practical necessity. Our EMVB implementation is specifically designed to mitigate the high I/O costs of this heavyweight paradigm.
\squishend

\subsection{Execution Module}
\label{subsec: execution module}
To minimize latency and support high-concurrency workloads, the framework implements a fine-grained, push-based execution model. Unlike traditional pull-based models where operators request data, data is actively pushed to the next operator as it becomes available. This approach reduces idle time and is highly effective for streaming search queries. Each operator in the pipeline is designed to execute concurrently. The framework manages a dedicated thread pool, strategically assigning operator tasks to available cores to maximize parallelism while mitigating resource contention. This combination of DAG-based planning and a concurrent, push-based engine allows the framework to efficiently coordinate complex, multi-path queries with minimal overhead.

\begin{figure}[!t]
\centering
\setlength{\abovecaptionskip}{0.1cm}
\setlength{\belowcaptionskip}{-0.3cm}
\includegraphics[width=0.85\linewidth]{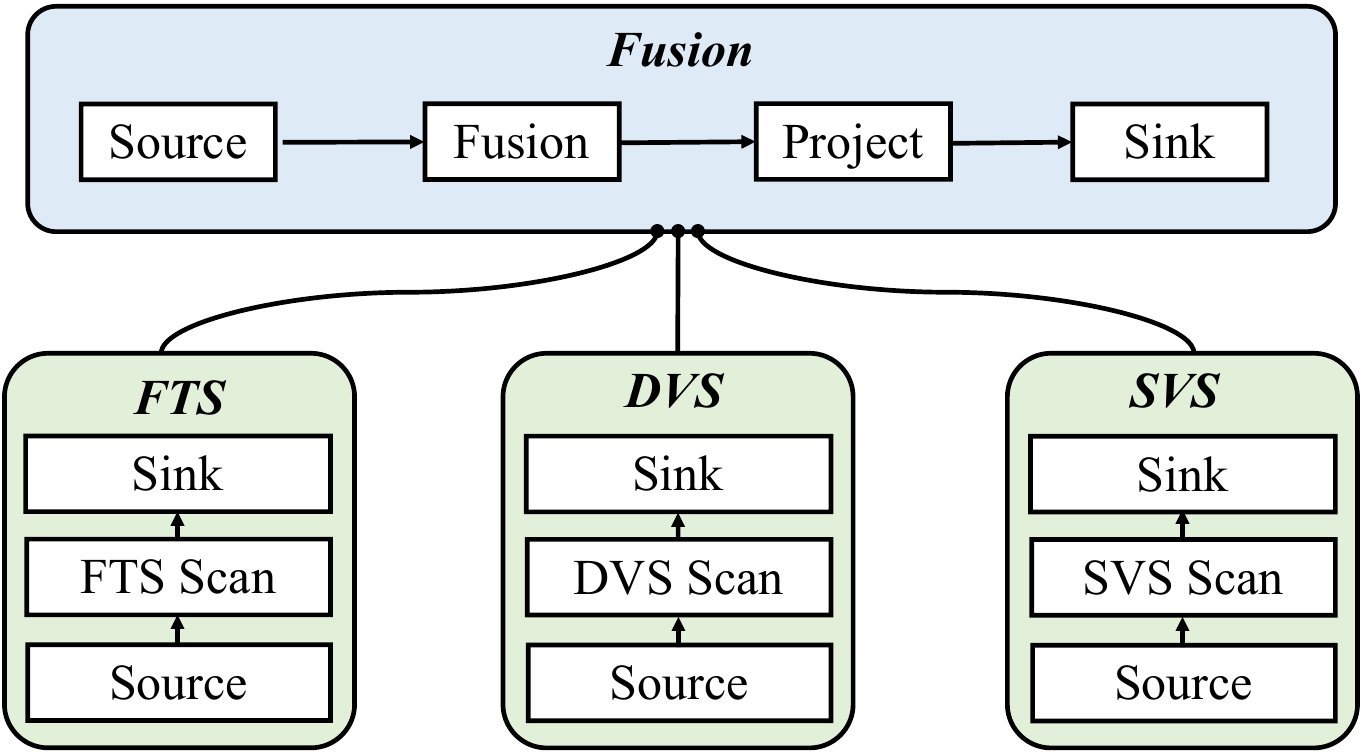}
\caption{Illustration of the pipelined DAG execution model for a three-path (\fts{}+\dvs{}+\svs{}) hybrid query.}
\label{fig:dag}
\end{figure}

\begin{myExa}
As illustrated in Figure~\ref{fig:dag}, a hybrid query is compiled into a query pipeline based on a Directed Acyclic Graph (DAG) of physical operators. The DAG structure enables parallel execution, where independent \texttt{Scan} operators for each retrieval path, such as an inverted index scan for \fts{} or a graph traversal for \dvs{}, run concurrently. As these operators generate candidates, results are streamed to a final \texttt{Fusion} operator. This operator incrementally applies a specified re-ranking strategy (e.g., RRF) to merge the incoming candidate lists. This pipelined model allows the fusion computation to overlap with the ongoing scan operations, minimizing end-to-end query latency.
\end{myExa}

\subsection{Algorithm Implementations}
\label{subsec: alg impl}
Our framework's modularity supports specialized, state-of-the-art implementations for each component.
We implemented our {\fts} engine from scratch for tight integration with our storage and execution models, inspired by production-grade designs like \texttt{Lucene}\footnote{\url{https://github.com/apache/lucene}} and \texttt{Tantivy}\footnote{\url{https://github.com/quickwit-oss/tantivy}}. It uses the BM25 ranking function, with query performance accelerated by a Block-Max WAND (BMW) \cite{DingS11} dynamic pruning algorithm. The implementation also supports configurable tokenizers \cite{tokenizer-standard,tokenizer-ik} and document ID reassignment to improve data locality and pruning efficiency.
For {\svs}, we implement Block-Max Pruning (BMP) \cite{BMP}, a highly-efficient dynamic pruning algorithm. Our implementation includes its core components of block filtering \cite{MalliaSS21} and candidate block evaluation, optimized with SIMD instructions \cite{DimopoulosNS13} and a hybrid inverted-forward index structure to accelerate top-$k$ selection.
The {\dvs} component uses a state-of-the-art HNSW graph index \cite{HNSW}. Our version further integrates Locally-adaptive Vector Quantization (LVQ) \cite{AguerrebereBHTW23}, a two-stage quantization process that significantly reduces the index footprint (minimizing disk I/Os) and accelerates distance computations during graph traversal.
To manage the high overhead of {\ts}, we implement the EMVB technique \cite{NardiniRV24}. This includes its key optimizations for efficiency: a bit-vector pre-filtering mechanism to quickly discard irrelevant documents, SIMD-accelerated centroid computations, and Product Quantization (PQ)~\cite{PQ} to compress the large number of residual token vectors.
Finally, the fusion operator supports three distinct strategies: the lightweight and widely-used Reciprocal Rank Fusion (RRF) and Weighted Sum (WS), alongside our Tensor-based Re-ranking Fusion (TRF), which uses the fine-grained MaxSim score (Equation (\ref{equ: maxSim})) for more accurate semantic re-ranking.

%% file: sections/4_eval_setup.tex
\section{Evaluation Setup}
\label{sec: eval setup}
This section details the experimental setup designed to answer the research questions posed in Section \ref{sec: intro}. We describe the core experimental components, define our evaluation metrics, and specify the implementation details and hardware environment.

\begin{table*}[th]
 \fontsize{7.5pt}{4mm}\selectfont
  \caption{Statistics of datasets. The last four columns denote the corpus size for each data type.}
  \vspace{-0.3cm}
  \label{tab:datasets}
  \setlength{\tabcolsep}{.008\linewidth}{
  \begin{tabular}{l|l|l|l|l|l|l|l|l|l}
    \hline
    \textbf{Dataset} & \textbf{Domain} & \textbf{Task} & \textbf{\#Corpus} & \textbf{\#Query} & \textbf{Avg. Length of Docs} & \textbf{Full-Text} & \textbf{Sparse}  & \textbf{Dense } & \textbf{Tensor } \\
    \hline
    \hline
    MSMA(en) \cite{dataset-ms} & Miscellaneous & Passage Retrieval & 8,841,823 & 43 & 56 & 2.9GB & 13GB & 36GB & 254GB \\
    \hline
    DBPE(en) \cite{dataset-db} & Wikipedia & Entity Retrieval & 4,635,922 & 400 & 50 &1.4GB &6.9GB &18.2GB &58GB \\
    \hline
    MCCN(zh) \cite{AIR-Bench} & News & Question Answering & 935,162 & 339 & 1,263 &2.3GB & 6.9GB&3.8GB &148GB \\
    \hline
    TOUC(en) \cite{dataset-touche} & Miscellaneous & Argument Retrieval & 382,545 & 49 & 292 &184MB &1.6GB &1.5GB &56GB \\
    \hline
    MLDR(zh) \cite{abs-2402-03216} & Wiki., Wudao & Long-Document Retrieval & 200,000 & 800 & 4,249 &3.1GB &5.2GB &791MB & 186GB\\
    \hline
    MLDR(en) \cite{abs-2402-03216} & Wikipedia & Long-Document Retrieval & 200,000 & 800 & 3,308 &3.1GB &4.4GB &791MB &94GB \\
    \hline
    TREC(en) \cite{dataset-tr} & Bio-Medical & Bio-Medical Information Retrieval & 171,332 & 50 & 161 & 184MB & 554MB & 688MB & 16GB \\
    \hline
    FIQA(en) \cite{dataset-fi} & Finance & Question Answering & 57,638 & 648 & 132 &43MB &137MB &232MB &4GB \\
    \hline
    CQAD(en) \cite{dataset-cq} & StackExchange & Duplicate-Question Retrieval & 40,221 & 1,570 & 129 &19MB &63MB &164MB & 1GB\\
    \hline
    SCID(en) \cite{dataset-scidocs} & Scientific & Citation Prediction & 25,657 & 1,000 & 176 &30MB &89MB &106MB &2.4GB \\
    \hline
    SCIF(en) \cite{dataset-scifact} & Scientific & Fact Checking & 5,183 & 809 & 214 &7.5MB &23MB &24MB &676MB \\
    \hline
  \end{tabular}
  }\vspace{-0.3cm}
\end{table*}

\subsection{Experimental Components}
This section details the core components—datasets, models, and methods—that constitute our experimental evaluation.

\subsubsection{Datasets}
To ensure generalizability, we evaluate all methods on 11 real-world datasets, chosen for variety in domain, task, and scale (Table \ref{tab:datasets}). Most are from the widely-used BEIR benchmark \cite{thakur2beir}, supplemented by two multilingual datasets from MTEB \cite{MTEB} and a Chinese news question answering dataset from AIR-Bench \cite{AIR-Bench} to enhance diversity. The collection covers 9 distinct tasks (e.g., fact-checking) across 7 domains, with corpus sizes from 5.2K to 8.8M. With the exception of MSMA(en), all datasets are out-of-domain. This provides a test of zero-shot retrieval performance, as the embedding models were not fine-tuned on them.

\subsubsection{Embedding Models}
We use mainstream models to generate \dvs{}, \svs{}, and \ts{} embeddings. For \dvs{} and \svs{}, we use \texttt{BGE-M3} \cite{abs-2402-03216}, which supports long inputs (up to 8,192 tokens) and concurrently produces high-quality dense and sparse representations. While \texttt{BGE-M3} can generate token tensors, its 1,024-dimensional embeddings are impractical for large-scale {\ts}; we observed embedding <50\% of the MLDR(en) dataset would require 1.1TB. To address this, we use the highly efficient \texttt{answerai-colbert-small-v1} \cite{answerai-colbert-small-v1} for English {\ts}. With 33M parameters (vs. \texttt{BGE-M3}’s 560M), it produces compact 96-dimensional representations. For Chinese datasets, we use \texttt{Jina-ColBERT-v2} \cite{abs-2408-16672}, selecting its most compact 64-dimensional variant for space efficiency. To validate our findings with a more recent model, we also conduct experiments using \texttt{NV-Embed-v2} \cite{Lee0XRSCP25}, a leading model from the MTEB leaderboard.

\subsubsection{Evaluated Methods}
Our evaluation covers four single-path retrieval methods ({\fts}, {\dvs}, {\svs}, and {\ts}), each implemented with a state-of-the-art algorithm (Section \ref{subsec: alg impl}). From these, we construct and evaluate all 11 possible hybrid combinations: all six two-path, all four three-path, and the single four-path configuration. For hybrid methods, we primarily evaluate three re-ranking strategies: the widely-used Reciprocal Rank Fusion (RRF) and Weighted Sum (WS), alongside our Tensor-based Rank Fusion (TRF), which uses tensor representations for a fine-grained re-ranking of candidates. To broaden our comparison, we also compare against two widely-used heavyweight learning-based re-rankers, \texttt{gte-multilingual-reranker-base} (GTE)~\cite{ZhangZLXDTLYXHZ24} and \texttt{bge-reranker-v2-m3} (BGE)~\cite{abs-2402-03216}, as external baselines.

\subsection{Evaluation Metrics}
We evaluate each method along three key dimensions: effectiveness, efficiency, and resource overhead.

\subsubsection{Effectiveness}
We measure retrieval effectiveness using Normalized Discounted Cumulative Gain at rank $k$ (nDCG@$k$), with $k=10$ by default. As our primary metric, nDCG@$k$ is standard in information retrieval (IR) benchmarks \cite{thakur2beir,abs-2402-14151,abs-2407-02883} due to its ability to handle graded relevance judgments and properly account for document rank. Theoretical analyses also confirm its comparability across methods \cite{WangWLHL13}. These properties make it more robust and informative than rank-unaware measures like Precision/Recall or those limited to binary relevance, such as Mean Reciprocal Rank (MRR) and Mean Average Precision (MAP) \cite{thakur2beir}. To confirm the robustness of our principal findings, we also report Recall@10 and MRR@10 in our analysis (Section~\ref{sec:effectiveness_results}).

\subsubsection{Efficiency and Overhead}
We evaluate system performance on both online query processing and offline index construction.
\textit{Query Performance}: We measure online efficiency via Query Latency (both average and P99 tail latency) and Throughput (Queries Per Second, QPS, under a multi-threaded batch workload).
\textit{Indexing Performance}: We measure offline efficiency via total Indexing Time (including data import and construction) and storage overhead via on-disk Index Size.
\textit{Resource Overhead}: We measure Peak Memory Consumption during both online querying and offline index construction. Our framework utilizes memory mapping (mmap) for index access, so the observed online memory footprint may be lower than the total index size.

\subsection{Implementation and Environment}
This section details the specific implementation parameters and the hardware environment used to conduct our experiments.

\subsubsection{Implementation Details}
All components are configured using parameters consistent with established best practices to ensure a fair comparison.
\textbf{\fts}: For the BM25 scoring function (refer to Equation (\ref{equ:bm25})), we use the standard default parameters of $k_1=1.2$ and $b=0.75$. We use the standard tokenizer \cite{tokenizer-standard} for English datasets and the IK tokenizer \cite{tokenizer-ik} for Chinese.
\textbf{\svs}: The Block-Max Pruning algorithm is configured with a block size of 8, in line with recommended settings for the method.
\textbf{\dvs}: Our HNSW index is constructed with a maximum neighbor size (\texttt{M}) of 16 and a candidate neighbor size (\texttt{efconstruction}) of 200, which are common settings in practice \cite{LiZAH20,HVS}. The LVQ enhancement is applied as detailed in its original work \cite{AguerrebereBHTW23}.
\textbf{\ts}: The EMVB algorithm is configured with 8192 centroids and 32 subspaces for its Product Quantization (PQ) stage, conforming to typical configurations for this technique \cite{AndreKS15,PQ,ZhanM0GZM21}.
\textbf{Re-ranking}: For RRF, the smoothing parameter $\kappa$ is set to 60, a widely-used value \cite{ChenZLBN22,CormackCB09}. For WS, the weight of each path is set to its nDCG@10 score from its single-path evaluation on the respective dataset. TRF re-ranks the candidate pool by computing the MaxSim score (Equation (\ref{equ: maxSim})) between the query's tokens and each document's pre-computed token tensors. For all hybrid methods, the candidate list size per path ($k_0$) is 10 by default, a value chosen based on initial analysis to balance effectiveness and efficiency (see Section~\ref{subsec: exp rq3}). Unless otherwise specified, RRF is the default re-ranking strategy.

\subsubsection{Experimental Environment}
All embedding generation was performed on two NVIDIA RTX 5000 Ada Generation GPUs. The core retrieval framework is implemented in C++ for performance, with a user-friendly Python API for accessibility. All experiments were conducted on a Linux server equipped with two Intel Xeon E5-2650 v4 CPUs (2.20 GHz). Each CPU comprises 12 cores and supports hyper-threading, providing 48 logical processors. The system is configured with 125GB of main memory (RAM). To ensure reliability, each experiment was executed at least three times, and the results were averaged to minimize environmental interference.


%% file: sections/5_results.tex
\begin{table*}[th]
 \fontsize{7.5pt}{4mm}\selectfont
  \caption{Retrieval accuracy (nDCG@10) of different methods with {\fts}, {\svs}, and {\dvs} across 11 datasets. In hybrid search, for each dataset column, the left values use RRF re-ranking, and the right values use TRF re-ranking. The best score on a given dataset is bolded, and the second-best is underlined. {\ts} is omitted due to high storage and computation costs.}
  \vspace{-0.3cm}
  \label{tab:accuracy hybrid search}
  \setlength{\tabcolsep}{.0042\linewidth}{
  \begin{tabular}{l|l|l|l|l|l|l|l|l|l|l|l|l|l|l|l|l|l|l|l|l|l|l|l}
    \hline
     \textbf{Method($\downarrow$)}& \textbf{Dataset($\to$)} & \multicolumn{2}{c|}{\textbf{MSMA(en)}} & \multicolumn{2}{c|}{\textbf{DBPE(en)}} & \multicolumn{2}{c|}{\textbf{MCCN(zh)}} & \multicolumn{2}{c|}{\textbf{TOUC(en)}} & \multicolumn{2}{c|}{\textbf{MLDR(zh)}} & \multicolumn{2}{c|}{\textbf{MLDR(en)}} & \multicolumn{2}{c|}{\textbf{TREC(en)}} & \multicolumn{2}{c|}{\textbf{\textbf{FIQA(en)}}} & \multicolumn{2}{c|}{\textbf{CQAD(en)}} & \multicolumn{2}{c|}{\textbf{SCID(en)}} & \multicolumn{2}{c}{\textbf{SCIF(en)}} \\
    \hline
    \hline
    \multirow{3}*{\textbf{Single-Path}} & {\fts} & \multicolumn{2}{c|}{.744} & \multicolumn{2}{c|}{.565} & \multicolumn{2}{c|}{.222} & \multicolumn{2}{c|}{.650} & \multicolumn{2}{c|}{.411} & \multicolumn{2}{c|}{.634} & \multicolumn{2}{c|}{.839} & \multicolumn{2}{c|}{.328} & \multicolumn{2}{c|}{.401} & \multicolumn{2}{c|}{.310} & \multicolumn{2}{c}{.704} \\
    \cline{2-24}
    ~ & {\svs} & \multicolumn{2}{c|}{.645} & \multicolumn{2}{c|}{.479} & \multicolumn{2}{c|}{.354} & \multicolumn{2}{c|}{.626} & \multicolumn{2}{c|}{.405} & \multicolumn{2}{c|}{.619} & \multicolumn{2}{c|}{.781} & \multicolumn{2}{c|}{.375} & \multicolumn{2}{c|}{.410} & \multicolumn{2}{c|}{.263} & \multicolumn{2}{c}{.647}  \\
    \cline{2-24}
    ~ & {\dvs} & \multicolumn{2}{c|}{.830} & \multicolumn{2}{c|}{.692} & \multicolumn{2}{c|}{.543} & \multicolumn{2}{c|}{.390} & \multicolumn{2}{c|}{.262} & \multicolumn{2}{c|}{.489} & \multicolumn{2}{c|}{.751} & \multicolumn{2}{c|}{\textbf{.532}} & \multicolumn{2}{c|}{.408} & \multicolumn{2}{c|}{.326} & \multicolumn{2}{c}{.715}\\
    \hline
    \hline
    \multirow{3}*{\textbf{Two-Path}} & {\fts}+{\svs} & .734 &  .867 & .608 & .678 & .328 & .438 & \textbf{.690} & \underline{.677} & .455 & \underline{.495} & .675 & \underline{.690} & .835 & .911 & .393 & .457 & .445 & .449 & .312 & .343 & .700 & .747 \\
    \cline{2-24}
    ~ & {\fts}+{\dvs} & .816 & \textbf{.894} & .668 & \textbf{.722} & .440 & \textbf{.624} & .604 & .618 & .411 & .471 & .646 & .680 & .832 & \underline{.920} & .471 & .508 & .463 & \underline{.465} & .347 & \textbf{.354} & .748 & \underline{.761} \\
    \cline{2-24}
    ~ & {\dvs}+{\svs} & .814 & .887 & .674 & .711 & .518 & \underline{.622} & .566 & .596 & .365 & .448 & .596 & .666 & .803 & .899 & .506 & \underline{.510} & .454 & .463 & .320 & \underline{.353} & .716 & .746  \\
    \hline
    \hline
    {\textbf{Three-Path}} & {\fts}+{\svs}+{\dvs} & \underline{.819} & \underline{.888} & .692 & \underline{.713} & .474 & \textbf{.624} & .654 & .627 & .451 & \textbf{.496} & .676 & \textbf{.693} & .865 & \textbf{.924} & .490 & .496 & \textbf{.486} & \underline{.465} & .345 & \underline{.353} & .739 & \textbf{.762}  \\
    \hline
  \end{tabular}
  }\vspace{-0.3cm}
\end{table*}

\begin{table*}[t!]
\centering
\caption{Effectiveness on the MLDR(en) dataset measured by Recall@10 and MRR@10.}
\vspace{-0.3cm}
\label{tab:extra_metrics}
\setlength{\tabcolsep}{.006\linewidth}{
\begin{tabular}{@{}lccc|cccc|cccc@{}}
\hline
\textbf{} & \multicolumn{3}{c|}{\textbf{Single-Path}} & \multicolumn{4}{c|}{\textbf{Multi-Path (RRF)}} & \multicolumn{4}{c}{\textbf{Multi-Path (TRF)}} \\ \cline{2-4} \cline{5-8} \cline{9-12} 
 & \textbf{\textsf{FTS}} & \textbf{\textsf{SVS}} & \textbf{\textsf{DVS}} & \textbf{\textsf{FTS}+\textsf{SVS}} & \textbf{\textsf{FTS}+\textsf{DVS}} & \textbf{\textsf{SVS}+\textsf{DVS}} & \textbf{\textsf{FTS}+\textsf{SVS}+DVS} & \textbf{\textsf{FTS}+\textsf{SVS}} & \textbf{\textsf{FTS}+\textsf{DVS}} & \textbf{\textsf{SVS}+\textsf{DVS}} & \textbf{\textsf{FTS}+\textsf{SVS}+\textsf{DVS}} \\ \hline
\textbf{Recall@10} & 0.755 & 0.735 & 0.614 & 0.794 & 0.766 & 0.739 & 0.798 & 0.806 & 0.790 & 0.771 & 0.814 \\
\textbf{MRR@10} & 0.595 & 0.582 & 0.450 & 0.637 & 0.608 & 0.552 & 0.637 & 0.652 & 0.643 & 0.629 & 0.653 \\ \hline
\end{tabular}
}
\end{table*}

\section{Experimental Results}
\label{sec: exp res}
This section presents our experimental evaluation, analyzing results across 11 real-world datasets to answer the research questions posed in Section~\ref{sec: intro}. We report the principal findings here; a more exhaustive set of results is available in~\cite{exp-ours}.

\subsection{Multi-Path Retrieval Effectiveness (RQ1)}
\label{sec:effectiveness_results}

\noindent\textbf{Finding 1}: While hybrid search accuracy generally improves with more retrieval paths, the overall performance is disproportionately constrained by the effectiveness of its weakest component.

The underlying principle of hybrid search is \textit{complementarity}: the assumption that fusing retrieval paths with different strengths will yield synergistic gains in accuracy.
Our results confirm this on many datasets. For instance, on the DBPE(en) dataset, combining an {\fts} path (nDCG@10 of 0.565) with an {\svs} path (0.479) under RRF improves the score to 0.608 (Table~\ref{tab:accuracy hybrid search}). However, this synergy is not guaranteed. A low-accuracy path can pollute the candidate pool with irrelevant documents, causing the hybrid system to underperform its best constituent path. We term this the \textbf{``weakest link''} phenomenon. This effect is demonstrated on the TOUC(en) dataset, where fusing a strong {\fts} path (0.650) with a weak {\dvs} path (0.390) degrades the final RRF score to 0.604.

\begin{figure}
  \setlength{\abovecaptionskip}{0cm}
  \setlength{\belowcaptionskip}{-0.4cm}
  \centering
  \footnotesize
  \stackunder[0.5pt]{\includegraphics[scale=0.40]{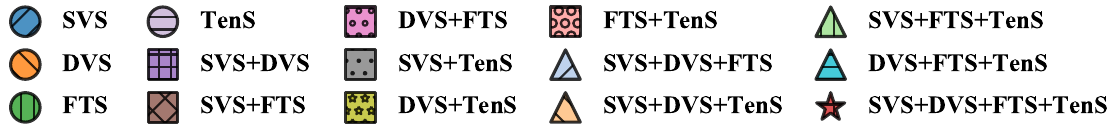}}{}
  \newline
  \stackunder[0.5pt]{\includegraphics[scale=0.25]{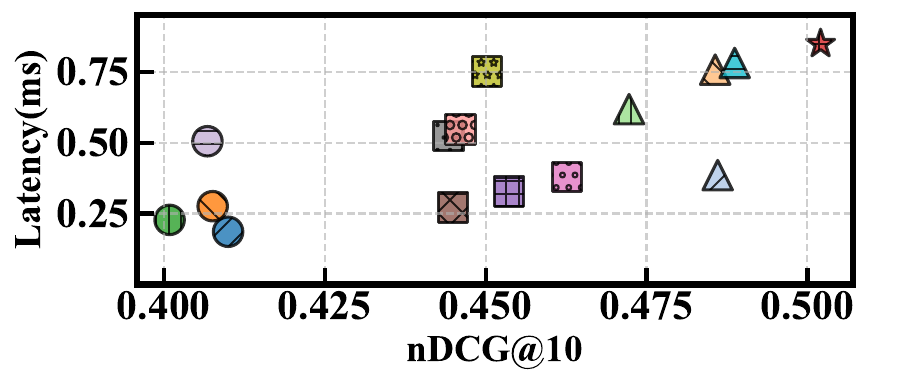}}{(a) CQAD(en)}
  \stackunder[0.5pt]{\includegraphics[scale=0.25]{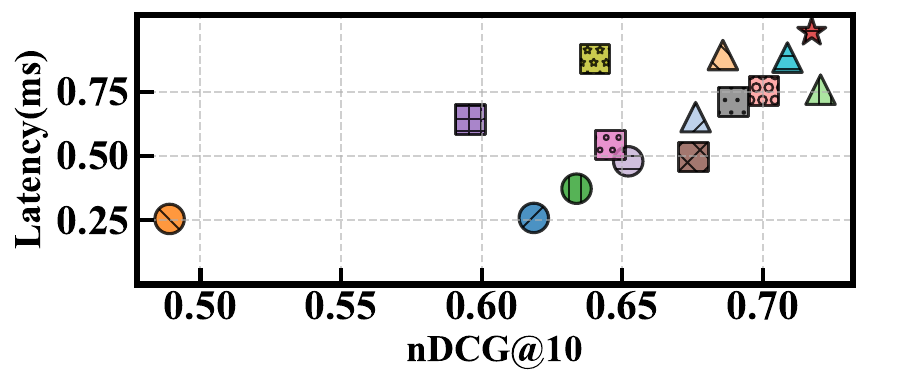}}{(b) MLDR(en)}
  \newline
  \stackunder[0.5pt]{\includegraphics[scale=0.25]{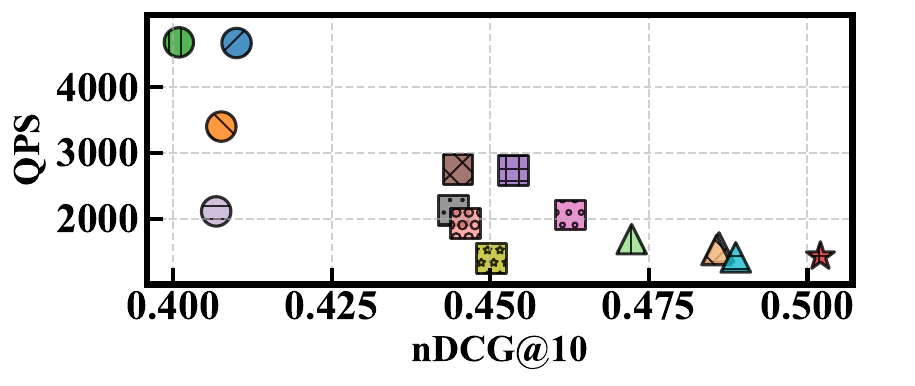}}{(c) CQAD(en)}
  \stackunder[0.5pt]{\includegraphics[scale=0.25]{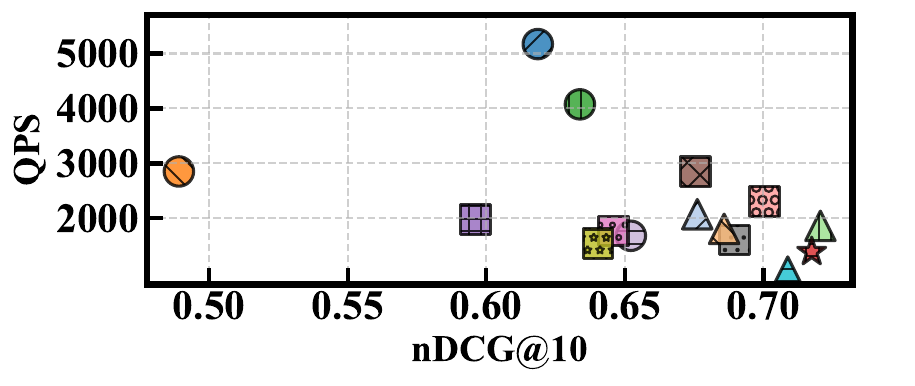}}{(d) MLDR(en)}
  \newline
  \caption{Accuracy vs. Efficiency of combination schemes.}
  \label{fig: accuracy vs efficiency trade-off v1}
\end{figure}

\subsubsection{Generality of the Phenomenon}
\begin{figure}
  \setlength{\abovecaptionskip}{0cm}
  \setlength{\belowcaptionskip}{-0.4cm}
  \centering
  \footnotesize
  \stackunder[0.5pt]{\includegraphics[scale=0.225]{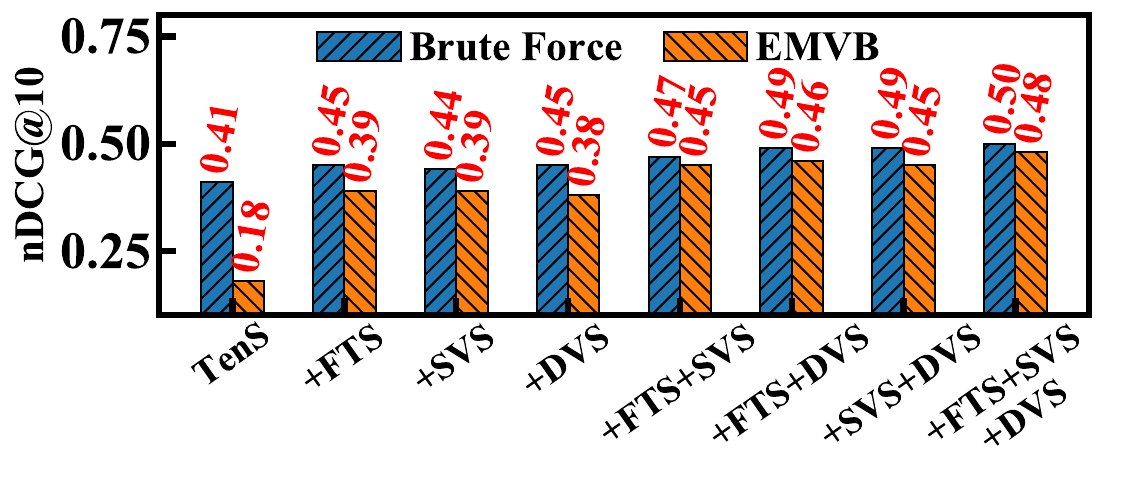}}{(a) CQAD(en)}
  \hspace{-0.15cm}
  \stackunder[0.5pt]{\includegraphics[scale=0.225]{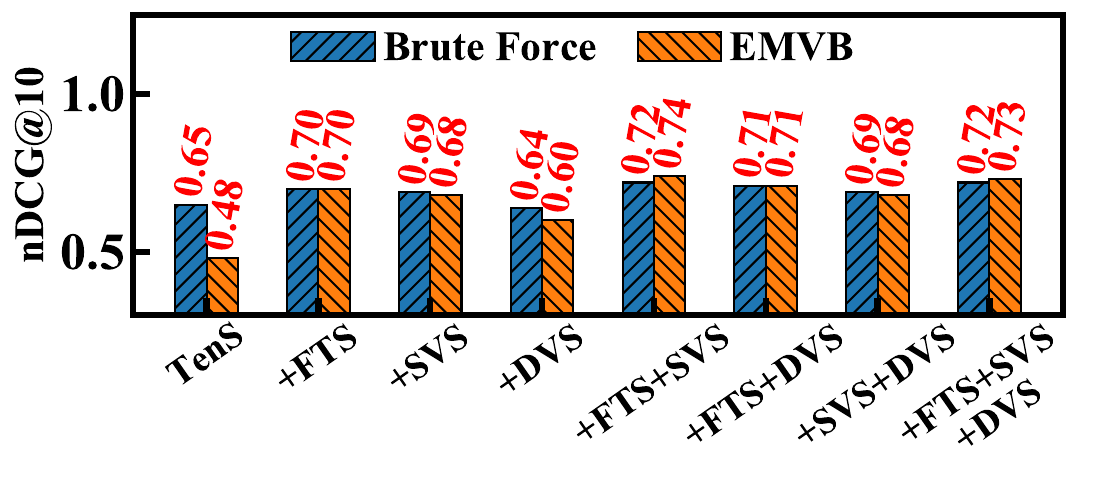}}{(b) MLDR(en)}
  \newline
  \caption{Accuracy comparison of brute-force and EMVB.}
  \label{fig: ts emvb}
\end{figure}

\begin{figure}
  \setlength{\abovecaptionskip}{0cm}
  \setlength{\belowcaptionskip}{-0.4cm}
  \centering
  \footnotesize
  \stackunder[0.5pt]{\includegraphics[scale=0.225]{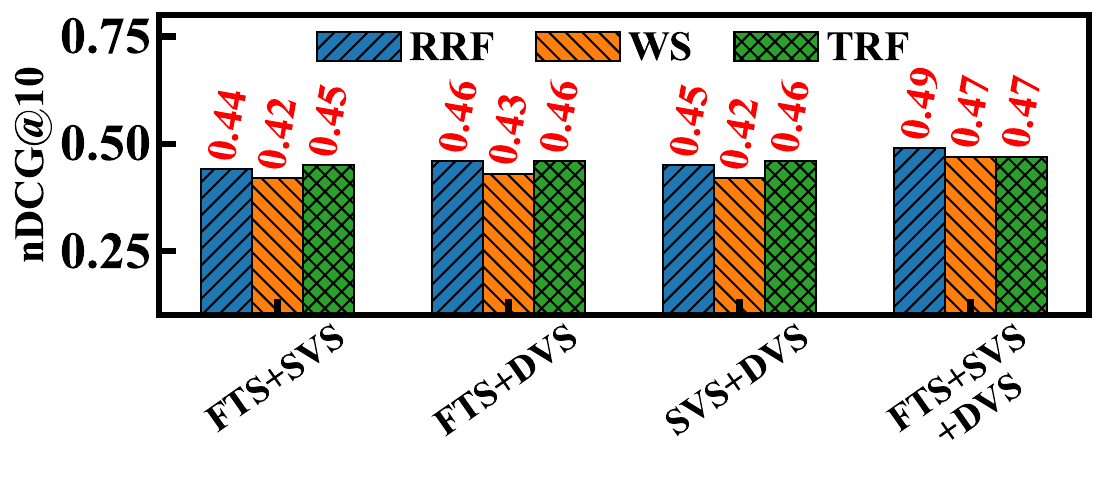}}{(a) CQAD(en)}
  \hspace{-0.15cm}
  \stackunder[0.5pt]{\includegraphics[scale=0.225]{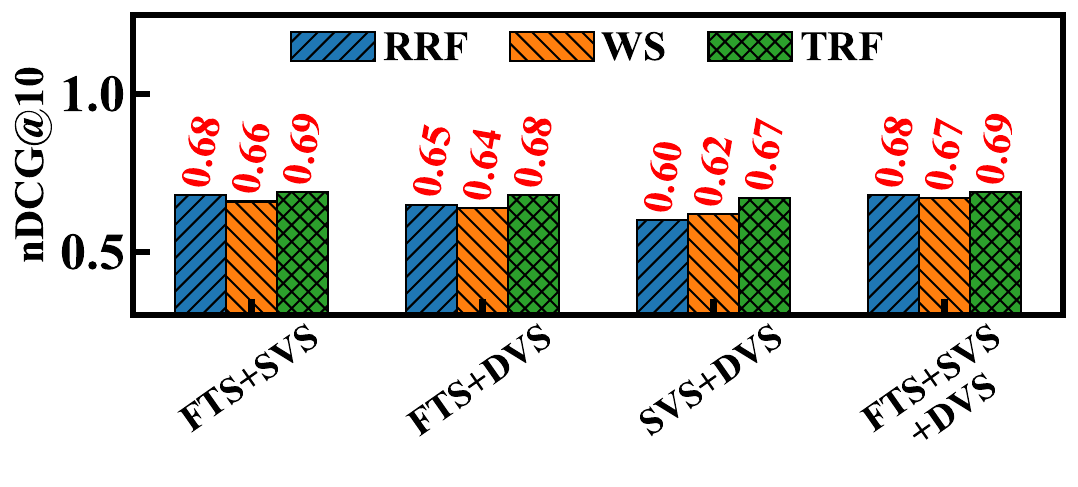}}{(b) MLDR(en)}
  \newline
  \caption{Accuracy comparison of RRF, WS, and TRF.}
  \label{fig: accuracy ranking}
\end{figure}

\begin{figure}
  \setlength{\abovecaptionskip}{0cm}
  \setlength{\belowcaptionskip}{-0.4cm}
  \centering
  \footnotesize
  \stackunder[0.5pt]{\includegraphics[scale=0.25]{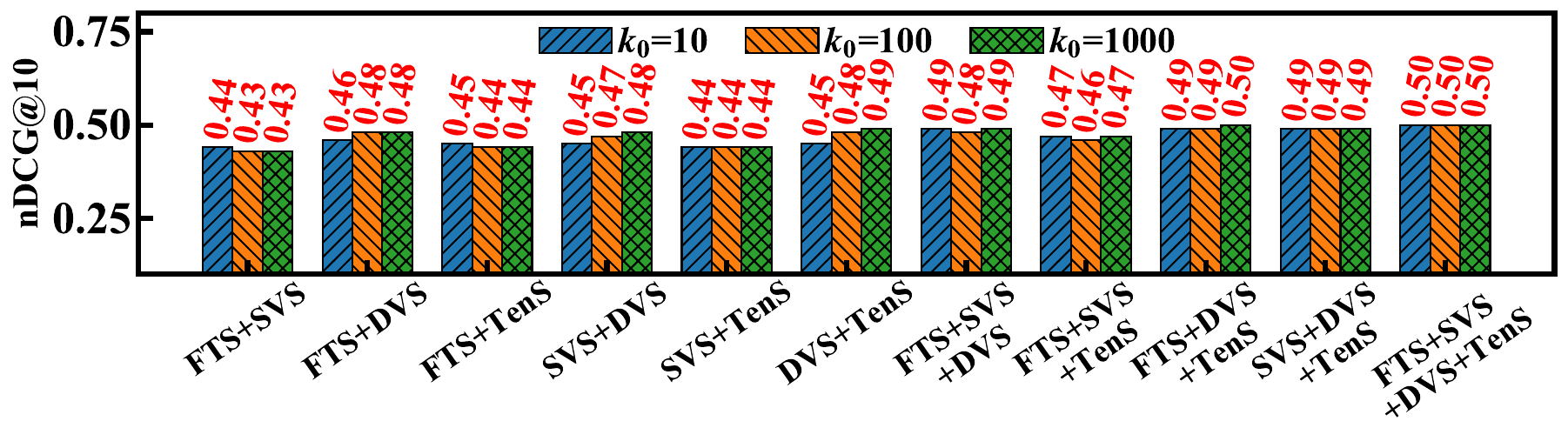}}{(a) CQAD(en)}
  \newline
  \stackunder[0.5pt]{\includegraphics[scale=0.25]{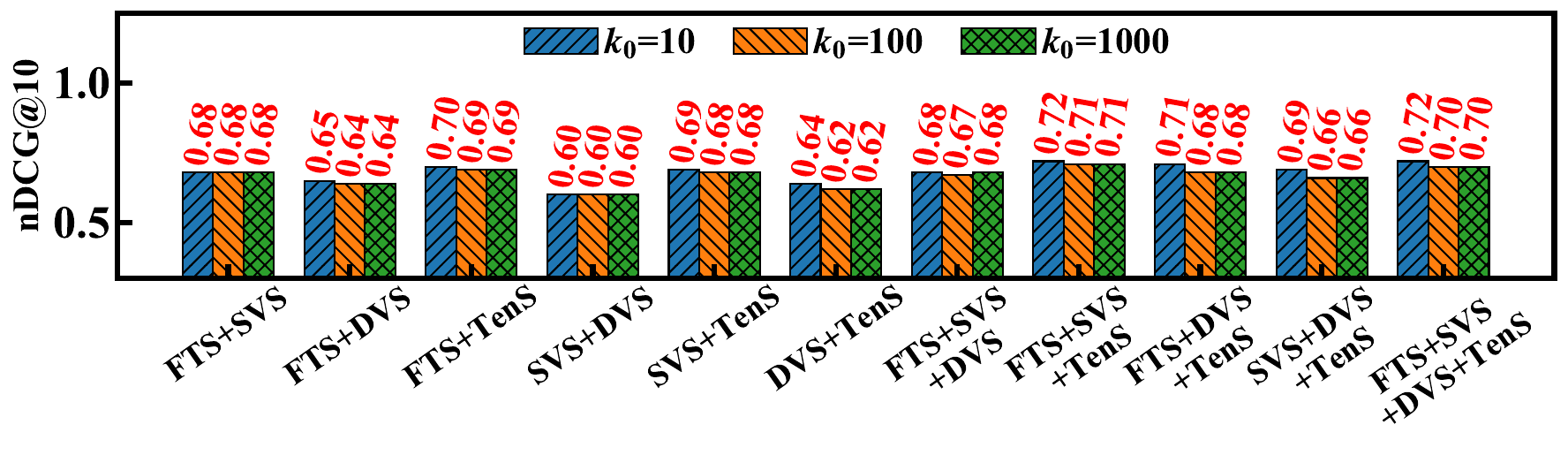}}{(b) MLDR(en)}
  \newline
  \caption{Accuracy under different numbers of candidates.}
  \label{fig: accuracy candi}
\end{figure}

To establish the generality of this finding, we verified its consistency across multiple experimental axes. The phenomenon is not an artifact of a single metric (Table~\ref{tab:extra_metrics}), persists across different embedding models (Table~\ref{tab:nv-embed-results}), and applies regardless of the specific algorithmic implementation (Figure~\ref{fig: ts emvb}).
Crucially, our results indicate that this degradation, originating in the candidate generation phase, is difficult to mitigate at the re-ranking stage. The effect holds true irrespective of the fusion strategy's complexity, from lightweight RRF (Figure~\ref{fig: accuracy ranking}) to heavyweight learning-based re-rankers (Table~\ref{tab:heavy_reranker_weak_link}). Moreover, simply increasing the candidate list size ($k_0$) fails to resolve the issue (Figure~\ref{fig: accuracy candi}). This underscores a key principle: the initial candidate quality imposes a hard accuracy ceiling, and once a weak path introduces irrelevant documents, the damage is difficult to reverse.

\subsubsection{Analysis of Influencing Factors}
We analyze the ``weakest link'' effect by two primary factors: the re-ranking mechanism and query characteristics. 
First, the failure mode depends on the re-ranking strategy. Rank-based methods like RRF are susceptible to high ranks from a weak path, irrespective of relevance. In contrast, TRF's semantic verification makes it vulnerable to ``hard negatives''---highly similar but irrelevant documents that achieve a high MaxSim score. Our case study in Section~\ref{sec:case-study} provides a concrete analysis of these distinct failure modes.
Second, which path becomes the ``weakest link'' is strongly correlated with query length. For datasets with long, descriptive queries such as MCCN(zh), the semantic {\dvs} path holds a substantial advantage over the lexical {\fts} path (nDCG@10 of 0.543 vs. 0.222). Conversely, for datasets with short keyword queries like TOUC(en), the trend reverses, with the precise {\fts} path significantly outperforming {\dvs} (0.650 vs. 0.390). This demonstrates a clear pattern: semantic paths thrive on the rich context provided by longer queries, whereas lexical paths excel at matching the exact keywords typical of shorter ones.

\begin{table*}[t!]
\centering
\caption{Effectiveness (nDCG@10) of hybrid search configurations with the \texttt{NV-Embed-v2} model.}
\vspace{-0.3cm}
\label{tab:nv-embed-results}
\setlength{\tabcolsep}{.004\linewidth}{
\begin{tabular}{@{}l|c|ccc|ccc@{}}
\hline
\multirow{2}{*}{\textbf{Dataset}} & \textbf{Single-Path} & \multicolumn{3}{c|}{\textbf{Multi-Path (RRF)}} & \multicolumn{3}{c}{\textbf{Multi-Path (TRF)}} \\ \cline{2-8} 
 & \textbf{\textsf{DVS}} & \textbf{\textsf{DVS}+\textsf{FTS}} & \textbf{\textsf{DVS}+\textsf{SVS}} & \textbf{All Three} & \textbf{\textsf{DVS}+\textsf{FTS}} & \textbf{\textsf{DVS}+\textsf{SVS}} & \textbf{All Three} \\ \hline
CQAD(en) & 0.481 & 0.537 & 0.564 & 0.547 & 0.506 & 0.505 & 0.486 \\
MLDR(en) & 0.489 & 0.511 & 0.427 & 0.616 & 0.662 & 0.641 & 0.686 \\ \hline
\end{tabular}
}
\end{table*}

\begin{table}[t!]
\centering
\caption{Effectiveness (nDCG@10) of heavyweight re-rankers on datasets with short (CQAD) and long (MLDR) documents.}
\vspace{-0.3cm}
\label{tab:heavy_reranker_weak_link}
\setlength{\tabcolsep}{.01\linewidth}{
\begin{tabular}{@{}lcccc@{}}
\hline
\multicolumn{5}{c}{\textbf{CQAD(en)}} \\
\cline{1-5}
\textbf{Methods} & \textbf{\textsf{FTS}+\textsf{SVS}} & \textbf{\textsf{FTS}+\textsf{DVS}} & \textbf{\textsf{DVS}+\textsf{SVS}} & \textbf{\textsf{FTS}+\textsf{SVS}+\textsf{DVS}} \\
\hline
\textbf{GTE} & 0.487 & 0.512 & 0.516 & 0.512 \\
\textbf{BGE} & 0.486 & 0.514 & 0.515 & 0.511 \\
\hline
\hline
\multicolumn{5}{c}{\textbf{MLDR(en)}} \\
\cline{1-5}
\textbf{Methods} & \textbf{\textsf{FTS}+\textsf{SVS}} & \textbf{\textsf{FTS}+\textsf{DVS}} & \textbf{\textsf{DVS}+\textsf{SVS}} & \textbf{\textsf{FTS}+\textsf{SVS}+\textsf{DVS}} \\
\hline
\textbf{GTE} & 0.471 & 0.454 & 0.466 & 0.438 \\
\textbf{BGE} & 0.488 & 0.472 & 0.480 & 0.456 \\
\hline
\end{tabular}
}
\vspace{-0.2cm}
\end{table}

\subsection{Combination Scheme Selection (RQ2)}
\label{subsec: exp rq2}

\noindent\textbf{Finding 2}: The optimal hybrid search architecture is a context-dependent trade-off among accuracy, query performance, and resource overhead; no single configuration excels universally.

We analyze the multi-dimensional trade-offs inherent in hybrid search architectures across three key dimensions: accuracy-latency, resource overhead, and context-specific factors.

\begin{table*}[t!]
\centering
\caption{Mean and P99 Latency (ms) comparison on the MLDR(en) dataset.}
\vspace{-0.3cm}
\label{tab:p99_latency}
\setlength{\tabcolsep}{.006\linewidth}{
\begin{tabular}{@{}lccc|cccc|cccc@{}}
\hline
\textbf{} & \multicolumn{3}{c|}{\textbf{Single-Path}} & \multicolumn{4}{c|}{\textbf{Multi-Path (RRF)}} & \multicolumn{4}{c}{\textbf{Multi-Path (TRF)}} \\ \cline{2-4} \cline{5-8} \cline{9-12} 
 & \textbf{\textsf{FTS}} & \textbf{\textsf{SVS}} & \textbf{\textsf{DVS}} & \textbf{\textsf{FTS}+\textsf{SVS}} & \textbf{\textsf{FTS}+\textsf{DVS}} & \textbf{\textsf{SVS}+\textsf{DVS}} & \textbf{\textsf{FTS}+\textsf{SVS}+DVS} & \textbf{\textsf{FTS}+\textsf{SVS}} & \textbf{\textsf{FTS}+\textsf{DVS}} & \textbf{\textsf{SVS}+\textsf{DVS}} & \textbf{\textsf{FTS}+\textsf{SVS}+\textsf{DVS}} \\ \hline
\textbf{Mean (ms)} & 0.37 & 0.26 & 0.26 & 0.50 & 0.54 & 0.64 & 0.65 & 0.65 & 0.73 & 0.72 & 0.87 \\
\textbf{P99 (ms)} & 0.53 & 0.45 & 0.63 & 0.58 & 0.85 & 0.83 & 0.91 & 0.77 & 0.95 & 1.00 & 1.01 \\ \hline
\end{tabular}
}
\vspace{-0.2cm}
\end{table*}

\subsubsection{Accuracy-Latency Trade-off}
\begin{figure}
  \setlength{\abovecaptionskip}{0cm}
  \setlength{\belowcaptionskip}{-0.4cm}
  \centering
  \footnotesize
  \stackunder[0.5pt]{\includegraphics[scale=0.36]{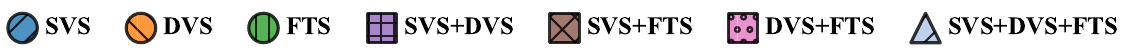}}{}
  \newline
  \stackunder[0.5pt]{\includegraphics[scale=0.28]{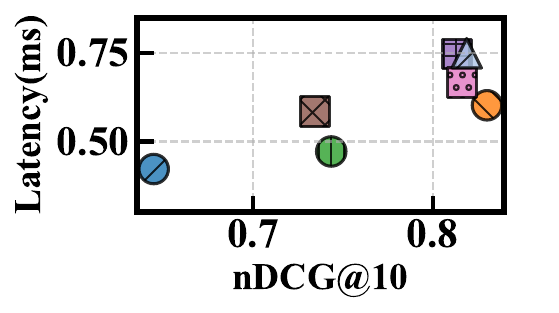}}{(a) MSMA(en)}
  \stackunder[0.5pt]{\includegraphics[scale=0.28]{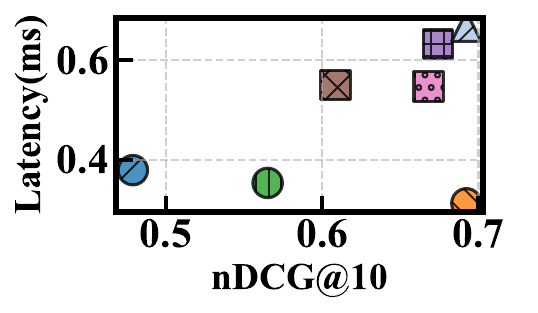}}{(b) DBPE(en)}
  \stackunder[0.5pt]{\includegraphics[scale=0.28]{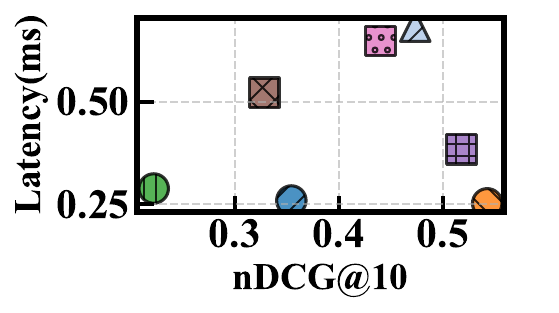}}{(c) MCCN(zh)}
  \newline
  \stackunder[0.5pt]{\includegraphics[scale=0.28]{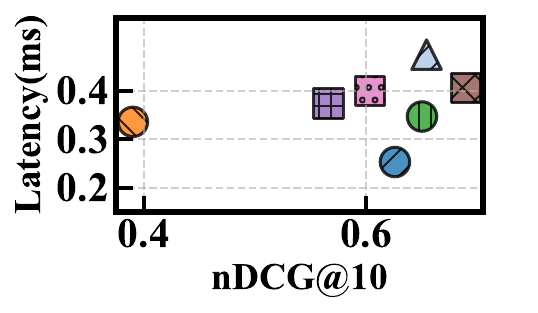}}{(d) TOUC(en)}
  \stackunder[0.5pt]{\includegraphics[scale=0.28]{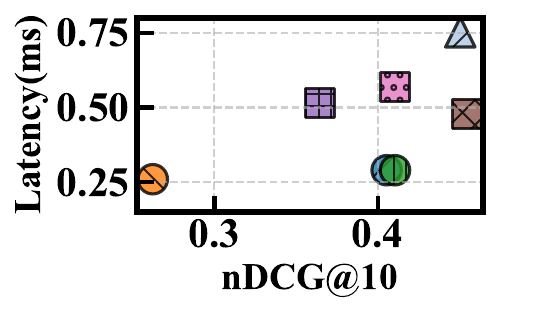}}{(e) MLDR(zh)}
  \stackunder[0.5pt]{\includegraphics[scale=0.28]{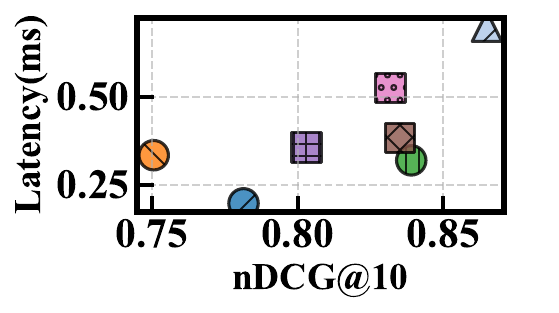}}{(f) TREC(en)}
  \newline
  \stackunder[0.5pt]{\includegraphics[scale=0.28]{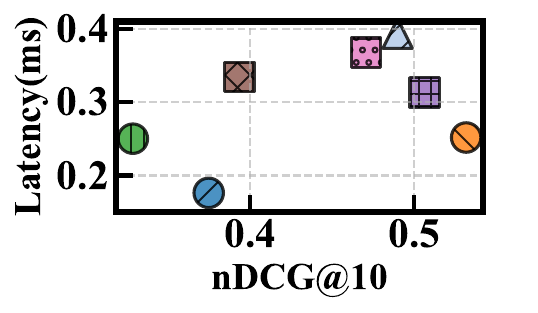}}{(g) FIQA(en)}
  \stackunder[0.5pt]{\includegraphics[scale=0.28]{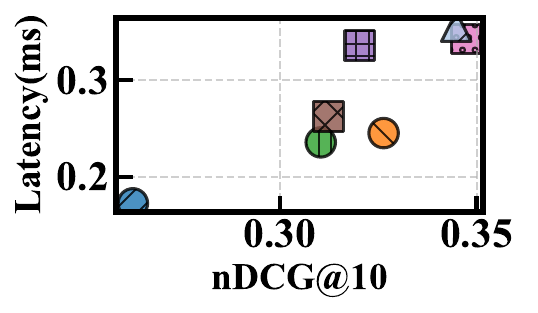}}{(h) SCID(en)}
  \stackunder[0.5pt]{\includegraphics[scale=0.28]{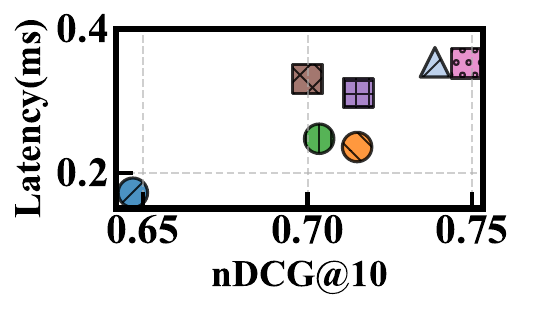}}{(i) SCIF(en)}
  \newline
  \caption{Cross-dataset latency-accuracy trade-offs.}
  \label{fig: ndcg-vs-latency v2}
\end{figure}

Adding retrieval paths improves accuracy at the cost of higher latency, a pattern consistent across datasets (Figures~\ref{fig: accuracy vs efficiency trade-off v1} and~\ref{fig: ndcg-vs-latency v2}). On CQAD(en), for instance, the four-path {\fts}+{\dvs}+{\svs}+{\ts} (red \textcolor{red}{$\star$}) achieves an nDCG@10 of 0.50, but its latency is 3.7$\times$ higher than the {\fts}-only path (green \textcolor{green}{$\bullet$}). This latency cost is inherent to the parallel execution model, where the final fusion step must await all path completions. Thus, a system's P99 tail latency is dictated by its slowest component (Table~\ref{tab:p99_latency}). For example, the {\fts}+{\dvs} combination inherits the {\dvs} path's high tail latency, making it substantially slower than {\fts} alone. This trade-off is complicated by query characteristics, as the slowest path can change: {\dvs} latency is constant, while {\fts} and {\svs} latency may increase with query length, demanding query-aware choices.

\subsubsection{Resource Overhead}
\begin{figure}
  \setlength{\abovecaptionskip}{0cm}
  \setlength{\belowcaptionskip}{-0.4cm}
  \centering
  \footnotesize
  \stackunder[0.5pt]{\includegraphics[scale=0.2]{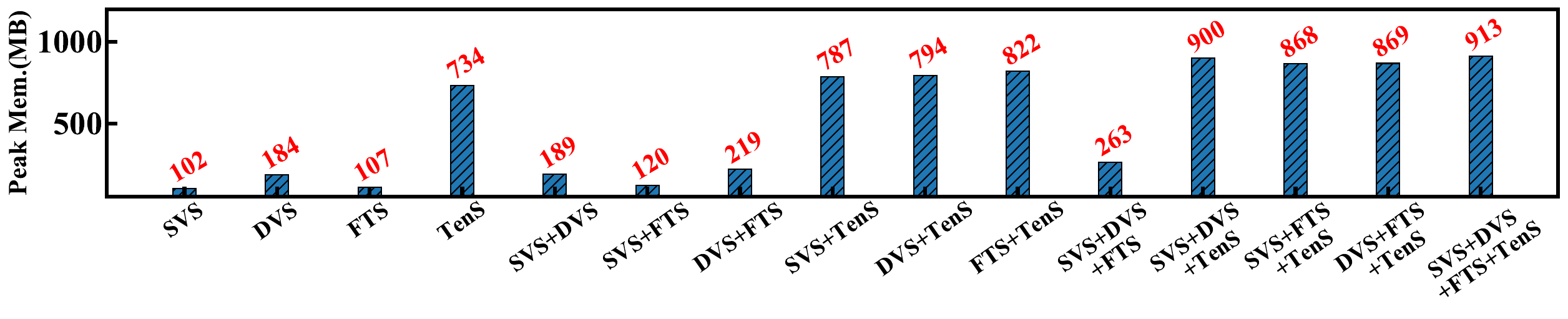}}{(a) CQAD(en)}
  \newline
  \stackunder[0.5pt]{\includegraphics[scale=0.2]{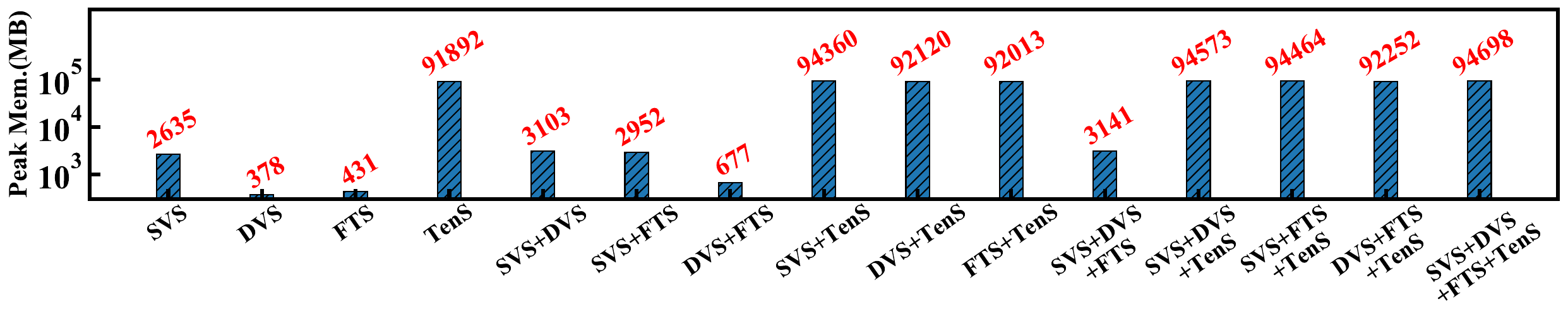}}{(b) MLDR(en)}
  \newline
  \caption{Memory consumption across retrieval methods.}
  \label{fig: mem v1}
\end{figure}

\begin{figure}
  \setlength{\abovecaptionskip}{0cm}
  \setlength{\belowcaptionskip}{-0.4cm}
  \centering
  \footnotesize
  \stackunder[0.5pt]{\includegraphics[scale=0.185]{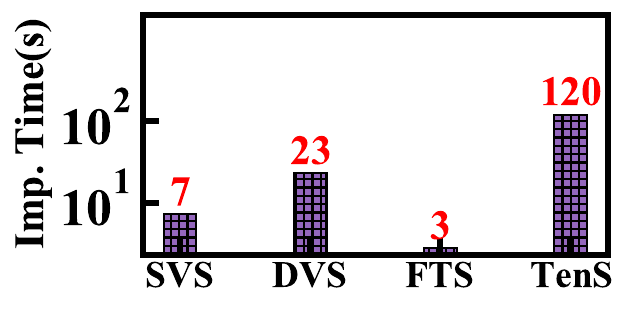}}{(a) CQAD(en)}
  \stackunder[0.5pt]{\includegraphics[scale=0.185]{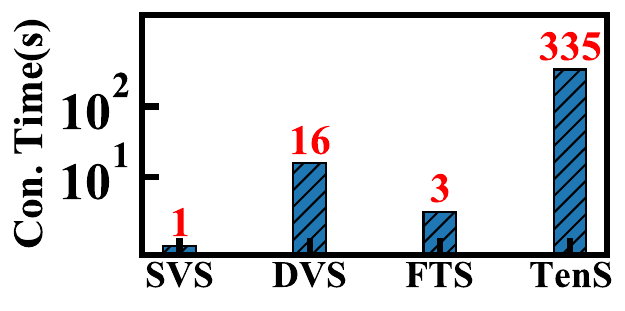}}{(b) CQAD(en)}
  \stackunder[0.5pt]{\includegraphics[scale=0.185]{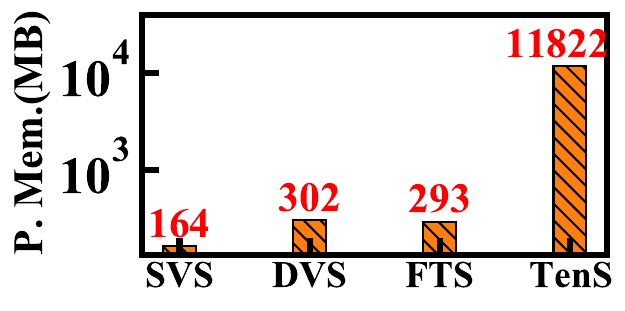}}{(c) CQAD(en)}
  \stackunder[0.5pt]{\includegraphics[scale=0.185]{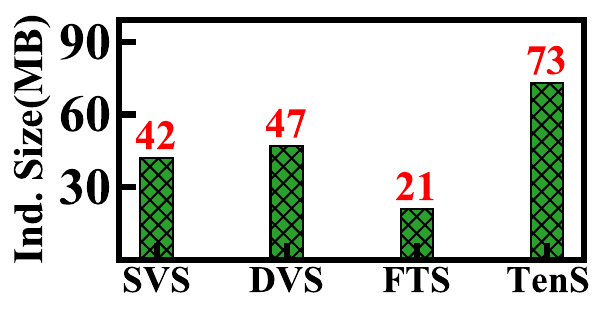}}{(d) CQAD(en)}
  \newline
  \stackunder[0.5pt]{\includegraphics[scale=0.185]{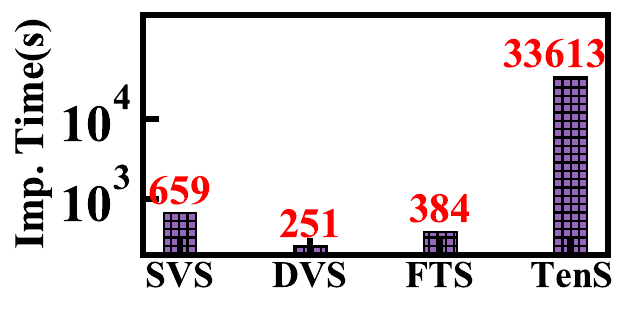}}{(e) MLDR(en)}
  \stackunder[0.5pt]{\includegraphics[scale=0.185]{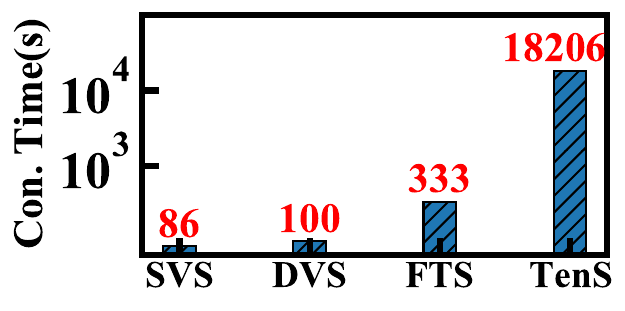}}{(f) MLDR(en)}
  \stackunder[0.5pt]{\includegraphics[scale=0.185]{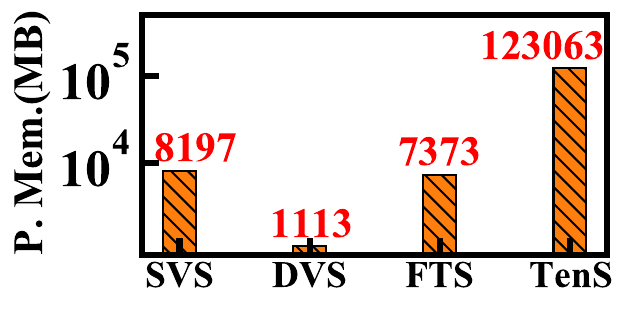}}{(g) MLDR(en)}
  \stackunder[0.5pt]{\includegraphics[scale=0.185]{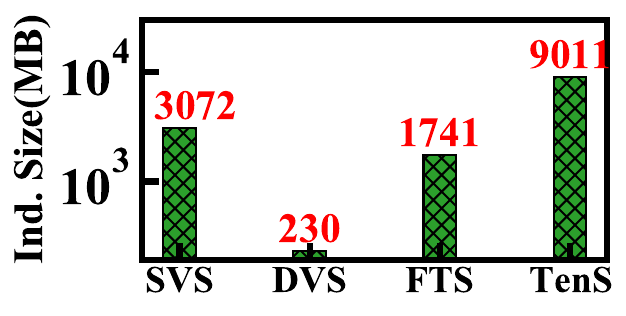}}{(h) MLDR(en)}
  \newline
  \caption{Indexing overhead of different retrieval paths.}
  \label{fig: index overhead}
\end{figure}

Hybrid search architectures introduce critical trade-offs in resource overhead, encompassing both online memory consumption and offline index construction. Resource usage generally scales with the number and type of paths combined, as shown in Figures~\ref{fig: mem v1} and~\ref{fig: index overhead}.
To illustrate the severity of this trade-off, we can examine the {\ts} paradigm as an example. Its resource cost is disproportionately high, with a memory footprint that can be orders of magnitude larger than other paradigms. This is a direct consequence of its per-token embedding representation; unlike {\dvs} where storage is constant per document, the storage footprint of {\ts} scales linearly with document length. This linear scaling is also reflected in its offline indexing costs, requiring substantially more time and peak memory to construct its index of token embeddings. These results demonstrate that the resource profile of each paradigm is a critical design consideration, as a single resource-intensive path can dominate a system's overall cost.

\subsubsection{Context-Dependent Trade-offs}
\begin{figure}
  \setlength{\abovecaptionskip}{0cm}
  \setlength{\belowcaptionskip}{-0.4cm}
  \centering
  \footnotesize
  \stackunder[0.5pt]{\includegraphics[scale=0.2]{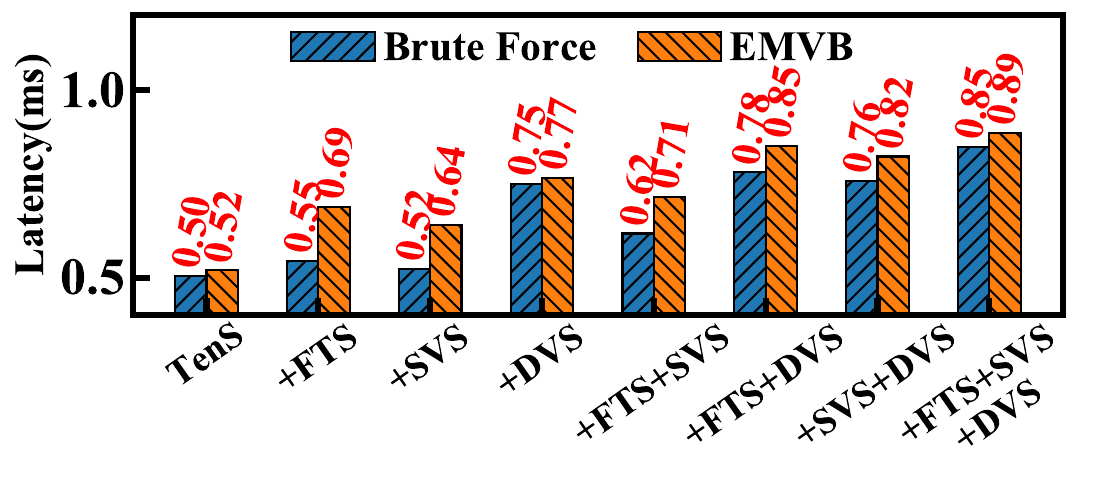}}{(a) CQAD(en)}
  \stackunder[0.5pt]{\includegraphics[scale=0.2]{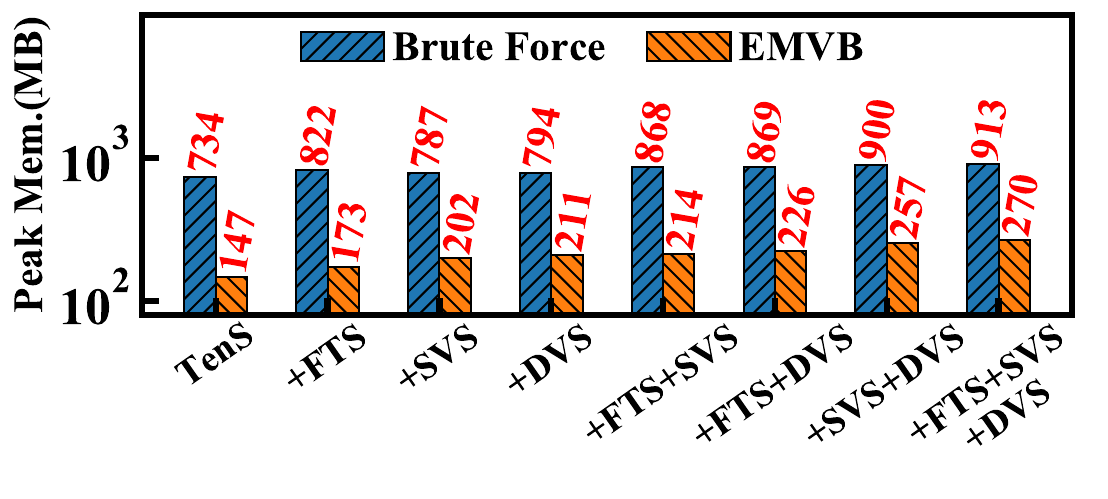}}{(b) CQAD(en)}
  \newline
  \stackunder[0.5pt]{\includegraphics[scale=0.2]{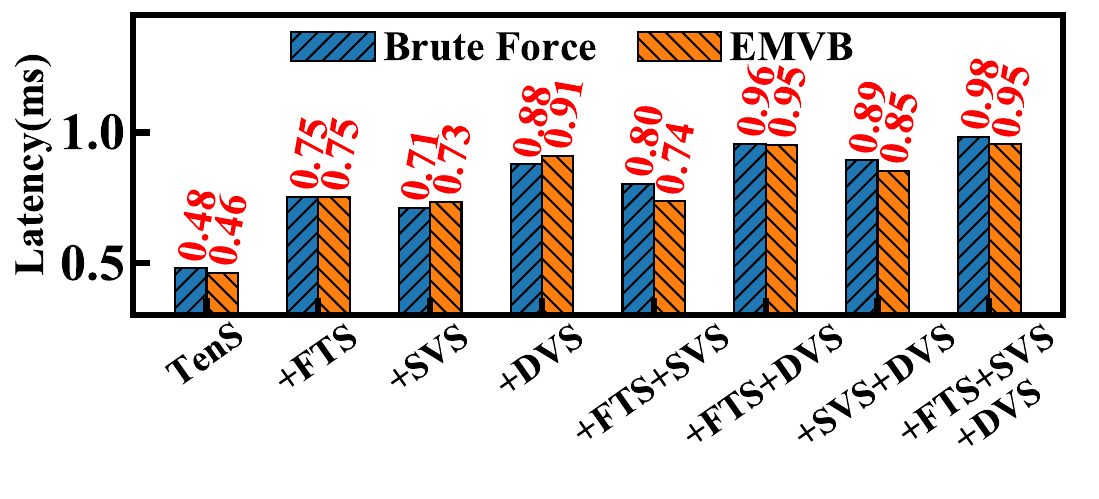}}{(c) MLDR(en)}
  \stackunder[0.5pt]{\includegraphics[scale=0.2]{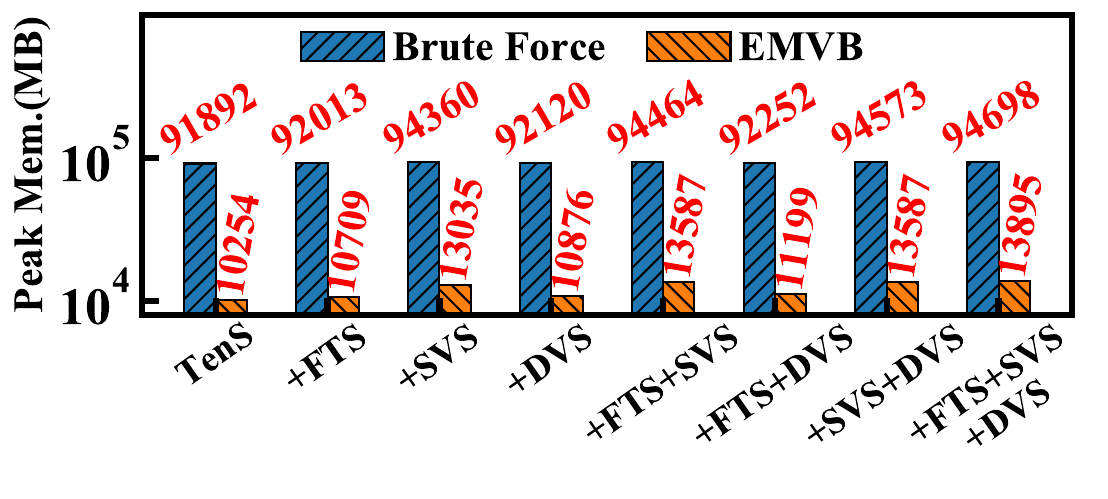}}{(d) MLDR(en)}
  \newline
  \caption{Online overhead of brute-fore and EMVB.}
  \label{fig: tensor overhead comp}
\end{figure}

The optimal trade-off point is not static but shifts based on algorithmic and data characteristics. For instance, an optimized algorithm like EMVB can dramatically reduce the resource cost of {\ts}, but at the expense of retrieval accuracy (Figure~\ref{fig: tensor overhead comp}). Similarly, data characteristics alter the balance. On corpora with long documents, the fixed-size vectors of {\dvs} are highly efficient to index, but may suffer from information loss; in contrast, the representations for {\fts} and {\svs} retain more term-level detail at the cost of larger index sizes. These factors reinforce our central finding: selecting a hybrid architecture is not a search for a single best configuration, but a complex optimization problem that must be tailored to specific algorithms, data characteristics, and application requirements.

\begin{figure}
  \setlength{\abovecaptionskip}{0cm}
  \setlength{\belowcaptionskip}{-0.4cm}
  \centering
  \footnotesize
  \stackunder[0.5pt]{\includegraphics[scale=0.225]{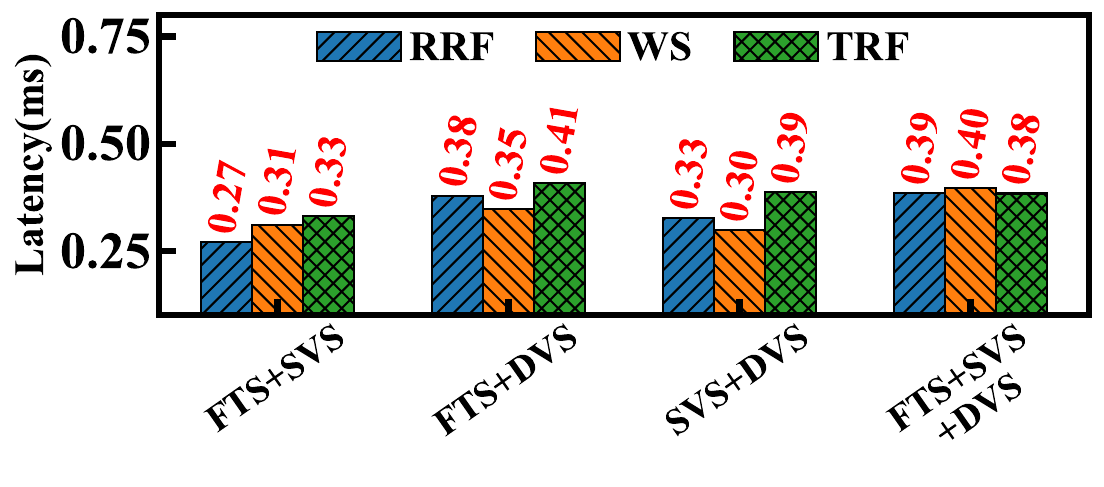}}{(a) CQAD(en)}
  \hspace{-0.15cm}
  \stackunder[0.5pt]{\includegraphics[scale=0.225]{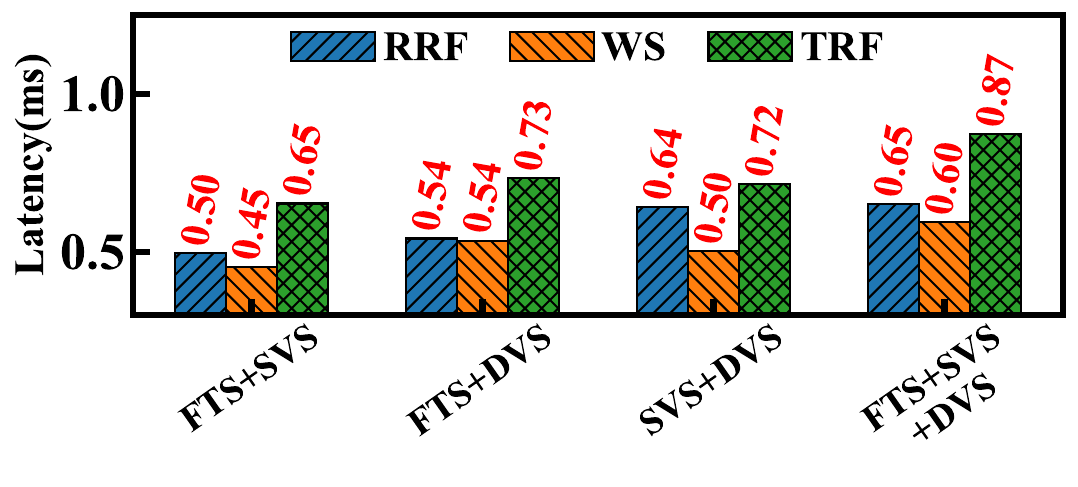}}{(b) MLDR(en)}
  \newline
  \caption{Retrieval latency under RRF, WS, and TRF.}
  \label{fig: latency ranking}
\end{figure}

\subsection{Cross-Path Re-ranking (RQ3)}
\label{subsec: exp rq3}

\noindent\textbf{Finding 3}: While lightweight fusion methods like RRF are efficient, TRF offers a path to significantly higher accuracy by leveraging fine-grained semantics at a fraction of the cost of a full {\fts} path.

Our experiments show that the choice of re-ranking strategy is critical to a hybrid system's final accuracy. We position TRF by comparing it against lightweight fusion, heavyweight learning-based re-rankers, and a full tensor retrieval path.

\subsubsection{TRF vs. Lightweight Fusion}TRF consistently demonstrates a significant accuracy advantage over widely-used lightweight methods like RRF and WS (Figure~\ref{fig: accuracy ranking}, Table~\ref{tab:accuracy hybrid search}). While RRF and WS operate on coarse document-level signals such as ranks or scores, TRF performs fine-grained semantic verification. It leverages the MaxSim computation to model token-to-token interactions between the query and each candidate. This effect is pronounced on the DBPE(en) dataset for the {\fts}+{\dvs} combination, where switching from RRF (nDCG@10 of 0.668) to TRF (0.722) yields an 8.1\% accuracy improvement, as we demonstrate with a query-level analysis in our case study (Section~\ref{sec:case-study}).

\subsubsection{TRF vs. Learning-based Re-rankers}
Compared to heavyweight learning-based re-rankers like GTE and BGE, TRF offers a more practical solution for in-database fusion. While heavyweight models excel on short-document datasets like CQAD(en), our results reveal a critical weakness on long-document corpora like MLDR(en), where their performance is substantially lower than even RRF (Table~\ref{tab:heavy_reranker_weak_link}). This failure stems from the fixed input token limits of cross-encoders (e.g., 512 tokens), which force document truncation and lead to critical information loss. Furthermore, these models incur over 100$\times$ more ranking time than TRF, rendering them impractical for the in-database fusion scenarios that are the focus of our work.

\subsubsection{TRF vs. TenS}
\begin{figure}
  \setlength{\abovecaptionskip}{0cm}
  \setlength{\belowcaptionskip}{-0.4cm}
  \centering
  \footnotesize
  \stackunder[0.5pt]{\includegraphics[scale=0.225]{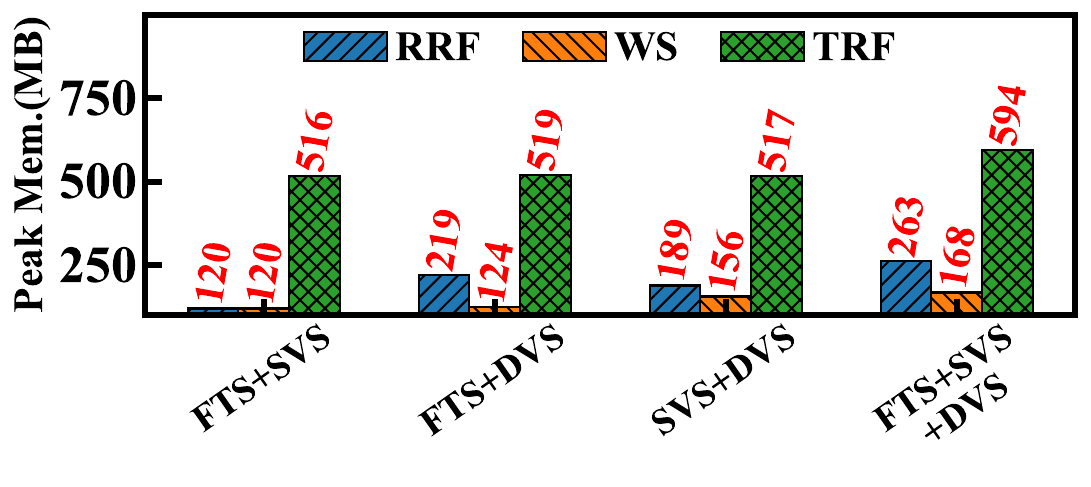}}{(a) CQAD(en)}
  \hspace{-0.15cm}
  \stackunder[0.5pt]{\includegraphics[scale=0.225]{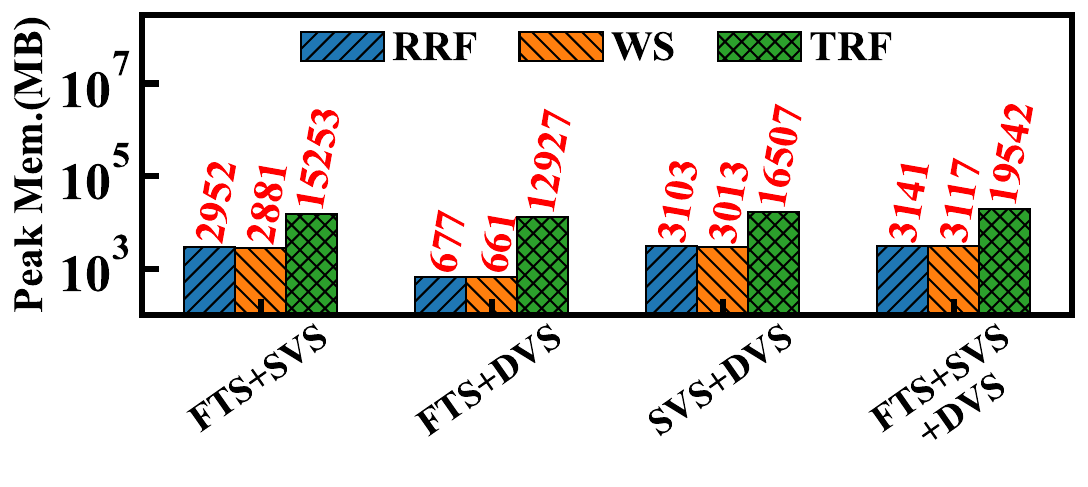}}{(b) MLDR(en)}
  \newline
  \caption{Memory overhead across re-ranking methods.}
  \label{fig: mem ranking}
\end{figure}

While TRF incurs higher latency and memory overhead than lightweight fusion (Figures~\ref{fig: latency ranking} and~\ref{fig: mem ranking}), its primary advantage lies in its cost-effectiveness compared to a full TenS path. The operational scope is the key difference: TRF re-ranks a small set of candidates, whereas a full TenS path must search a massive index. This architectural design yields dramatic resource savings; for example, on MLDR(en), using TRF on the {\fts}+{\dvs} candidate set reduces peak memory usage by 86\% compared to adding a full {\fts}+{\dvs}+{\ts} combination. This highlights a key insight: late interaction is more efficiently harnessed as a re-ranking mechanism on a pre-filtered candidate set than as a primary retrieval method.

\subsubsection{Sensitivity to Candidate Set Size ($k_0$)}
\begin{figure}
  \setlength{\abovecaptionskip}{-0.2cm}
  \setlength{\belowcaptionskip}{-0.4cm}
  \centering
  \footnotesize
  \stackunder[0.5pt]{\includegraphics[scale=0.25]{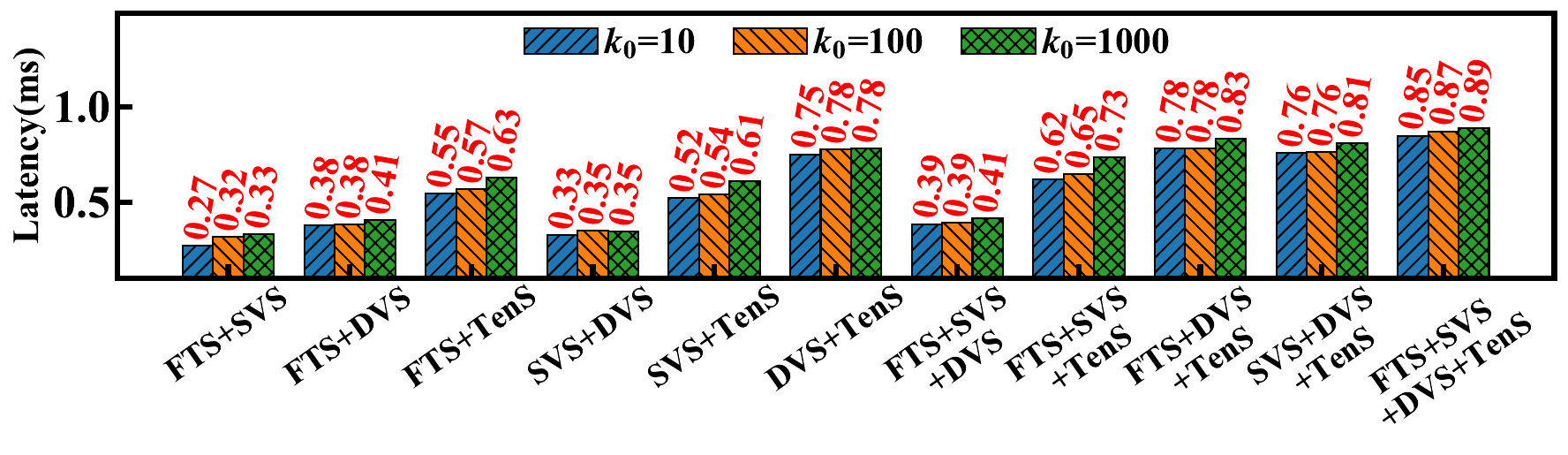}}{}
  \caption{Latency under different numbers of candidates.}
  \label{fig: latency candi}
\end{figure}

\begin{figure}
  \setlength{\abovecaptionskip}{-0.2cm}
  \setlength{\belowcaptionskip}{-0.4cm}
  \centering
  \footnotesize
  \stackunder[0.5pt]{\includegraphics[scale=0.25]{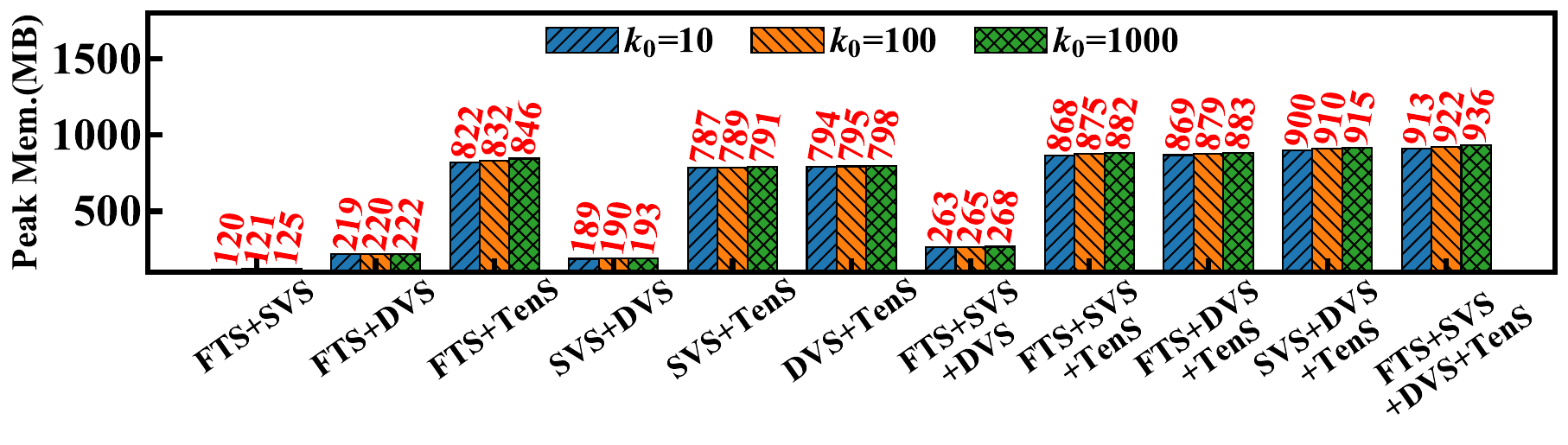}}{}
  \caption{Memory under different numbers of candidates.}
  \label{fig: mem candi}
  \vspace{-0.1cm}
\end{figure}

The performance overhead of TRF is a direct function of the number of candidates ($k_0$) being re-ranked, which provides a clear tuning parameter. Our analysis demonstrates that increasing $k_0$ yields diminishing returns for accuracy (Figure \ref{fig: accuracy candi}) while incurring a near-linear increase in latency and memory for TRF (Figures \ref{fig: latency candi} and \ref{fig: mem candi}). This allows system designers to directly balance accuracy requirements against latency and memory budgets by adjusting the candidate set size.

\begin{table*}[t]
\centering
\caption{Detailed Top-10 results for query \texttt{INEX\_LD-2009061} ("france second world war normandy") on the DBPE(en) dataset. Results show shortened document titles (e.g., \texttt{<...Normandy\_Campaign>}). Document relevance from qrels is shown in parentheses (Rel: 1 or 0). Bolded documents are relevant.}
\vspace{-0.3cm}
\label{tab:case-study}
\resizebox{\textwidth}{!}{%
\begin{tabular}{@{}l|ll|ll@{}}
\toprule
\textbf{Rank} & \multicolumn{1}{c}{\textbf{\fts} (nDCG@10: 0.624)} & \multicolumn{1}{c|}{\textbf{\dvs} (nDCG@10: 0.784)} & \multicolumn{1}{c}{\textbf{\fts+\dvs} (RRF, nDCG@10: 0.604)} & \multicolumn{1}{c}{\textbf{\fts+\dvs} (TRF, nDCG@10: 0.880)} \\ \midrule
1 & <...Bayeux...Cemetery> (0) & \textbf{<...military...Normandy...>} (1) & <...American\_Cemetery...> (0) & \textbf{<...Normandy\_Campaign>} (1) \\
2 & \textbf{<...Hillman\_Site>} (1) & <...American\_Cemetery...> (0) & <...Bayeux...Cemetery> (0) & \textbf{<...Invasion\_of\_Normandy>} (1) \\
3 & <...American\_Cemetery...> (0) & <...Operation\_Newton> (0) & \textbf{<...military...Normandy...>} (1) & <...American\_Cemetery...> (0) \\
4 & <...Order\_of\_battle...Epsom> (0) & <...Orglandes...cemetery> (0) & \textbf{<...Hillman\_Site>} (1) & <...Normandy> (0) \\
5 & \textbf{<...Bény-sur-Mer...War...>} (1) & \textbf{<...Falaise\_pocket>} (1) & <...Operation\_Newton> (0) & <...Order\_of\_battle...Epsom> (0) \\
6 & <...Duke\_of\_Normandy\_(pigeon)> (0) & <...Normandie-Niemen> (0) & <...Order\_of\_battle...Epsom> (0) & <...Bayeux...Cemetery> (0) \\
7 & <...Operation\_Cobra...battle> (0) & \textbf{<...Normandy\_Campaign>} (1) & <...Orglandes...cemetery> (0) & \textbf{<...military...Normandy...>} (1) \\
8 & <...Ranger\_Assault\_Group> (0) & <...Normandy> (0) & \textbf{<...Bény-sur-Mer...War...>} (1) & <...Operation\_Cobra...battle> (0) \\
9 & <...Saint-Jean-de-Daye\_Airfield> (0) & <...Order\_of\_battle...airborne...> (0) & \textbf{<...Falaise\_pocket>} (1) & <...Orglandes...cemetery> (0) \\
10 & <...Second\_Army\_(UK)> (0) & \textbf{<...Invasion\_of\_Normandy>} (1) & <...Duke\_of\_Normandy\_(pigeon)> (0) & \textbf{<...Falaise\_pocket>} (1) \\ \bottomrule
\end{tabular}
}
\vspace{-0.2cm}
\end{table*}

\subsection{Query-level Case Study}
\label{sec:case-study}

We analyze query \texttt{INEX\_LD-2009061} (``france second world war normandy'') from the DBPE(en) dataset to illustrate our findings, with results detailed in Table \ref{tab:case-study}. The query's intent (military events) is lexically ambiguous, with terms that also relate to memorials.

The RRF fusion of {\fts} and {\dvs} yields an nDCG@10 of 0.604, a score lower than the 0.784 achieved by the stronger {\dvs} path alone. This degradation exemplifies the ``weakest link'' phenomenon and stems from RRF's rank-based heuristic. As Table \ref{tab:case-study} shows, {\fts} ranks the irrelevant \texttt{<...Bayeux\_Cemetery>} first (lexical matches), while {\dvs} ranks the irrelevant \texttt{<...American\_Cemetery>} second (semantic score). RRF misinterprets the high ranks from both paths as agreement and incorrectly promotes these two irrelevant documents. The {\fts} path thus acts as a ``weakest link'', introducing high-ranking noise that degrades the final result.

In contrast, TRF re-scores the same candidate set and achieves a significantly higher nDCG@10 of 0.880. Its effectiveness stems from its fine-grained semantic verification using the MaxSim score. This token-level computation allows TRF to disambiguate the query's intent. Consequently, it demotes irrelevant documents like the cemeteries, as their constituent tokens (e.g., ``cemetery'', ``memorial'') are a poor semantic match for the query's military intent. Conversely, TRF elevates relevant documents such as \texttt{<...Normandy\_Campaign>} and \texttt{<...Invasion\_of\_Normandy>}, which {\dvs} had ranked only 7th and 10th. It promotes them because their tokens (e.g., ``campaign'', ``invasion'', ``battle'') align precisely with the query's semantics.
The TRF method is not flawless, as it still places the irrelevant \texttt{<...American\_Cemetery>} at rank 3, treating it as a hard negative. Nonetheless, this example highlights the difference in principles: RRF's reliance on rank-based fusion makes it susceptible to noise from a single weak path, whereas TRF's token-level verification provides greater resilience against such rank interference.

%% file: sections/6_discussion.tex
\section{Discussion and Future Work}
\label{sec: discussion}
Building upon the extensive experimental evaluation, this section synthesizes our findings and analyzes their implications.

\newcommand{\scorebar}[1]{
    \begingroup
    \color{black}\rule{#1\dimexpr0.4em\relax}{1.5ex}
    \color{gray!40}\rule{\dimexpr2em - #1\dimexpr0.4em\relax}{1.5ex}
    \endgroup
}

\begin{table}[t]
\centering
\fontsize{7.5pt}{9pt}\selectfont 
\caption{A qualitative summary of hybrid search architecture trade-offs. 
Ratings are shown as a 5-unit bar (e.g., \scorebar{4}), where more black indicates better performance (higher accuracy or higher efficiency/lower cost). 
We use \textsf{F}, \textsf{S}, \textsf{D}, and \textsf{T} as abbreviations for \textsf{FTS}, \textsf{SVS}, \textsf{DVS}, and \textsf{TenS}, respectively.
}
\vspace{-0.3cm}
\label{tab:qualitative_summary}
\setlength{\tabcolsep}{4.5pt}{
\begin{tabular}{@{}l l c c c c@{}}
\toprule
\textbf{\#Paths} & \textbf{Type} & \textbf{Accuracy} & \textbf{Efficiency} & \textbf{Memory} & \textbf{Indexing} \\
\midrule
\multirow{4}*{\textbf{1}} & {\textsf{F}} & \scorebar{2} & \scorebar{5} & \scorebar{5} & \scorebar{5} \\
& {\textsf{S}} & \scorebar{2} & \scorebar{5} & \scorebar{5} & \scorebar{5} \\
& {\textsf{D}} & \scorebar{2} & \scorebar{5} & \scorebar{5} & \scorebar{5} \\
& {\textsf{T}} & \scorebar{3} & \scorebar{2} & \scorebar{2} & \scorebar{1} \\
\hline
\multirow{6}*{\textbf{2}} & {\textsf{F}}+{\textsf{S}} & \scorebar{3} & \scorebar{4} & \scorebar{4} & \scorebar{4} \\
& {\textsf{F}}+{\textsf{D}} & \scorebar{3} & \scorebar{4} & \scorebar{4} & \scorebar{4} \\
& {\textsf{F}}+{\textsf{T}} & \scorebar{4} & \scorebar{2} & \scorebar{2} & \scorebar{1} \\
& {\textsf{S}}+{\textsf{D}} & \scorebar{3} & \scorebar{4} & \scorebar{4} & \scorebar{4} \\
& {\textsf{S}}+{\textsf{T}} & \scorebar{4} & \scorebar{2} & \scorebar{2} & \scorebar{1} \\
& {\textsf{D}}+{\textsf{T}} & \scorebar{4} & \scorebar{2} & \scorebar{2} & \scorebar{1} \\
\hline
\multirow{4}*{\textbf{3}}& {\textsf{F}}+{\textsf{S}}+{\textsf{D}} & \scorebar{4} & \scorebar{3} & \scorebar{3} & \scorebar{3} \\
& {\textsf{F}}+{\textsf{S}}+{\textsf{T}} & \scorebar{5} & \scorebar{2} & \scorebar{2} & \scorebar{1} \\
& {\textsf{F}}+{\textsf{D}}+{\textsf{T}} & \scorebar{5} & \scorebar{2} & \scorebar{2} & \scorebar{1} \\
& {\textsf{S}}+{\textsf{D}}+{\textsf{T}} & \scorebar{5} & \scorebar{2} & \scorebar{2} & \scorebar{1} \\
\hline
\textbf{4} & {\textsf{F}}+{\textsf{S}}+{\textsf{D}}+{\textsf{T}} & \scorebar{5} & \scorebar{1} & \scorebar{2} & \scorebar{1} \\
\bottomrule
\end{tabular}
}\vspace{-0.3cm}
\end{table}

\subsection{Principal Findings and Their Implications}
We now discuss our three principal findings and their practical implications. Table \ref{tab:qualitative_summary} summarizes the trade-offs of hybrid search architectures qualitatively.

\subsubsection{The ``Weakest Link'' Effect: The Critical Role of Path Quality}
Table \ref{tab:qualitative_summary} suggests that more paths often lead to higher accuracy. However, our results also show a hybrid system's accuracy is constrained by its least effective component, sometimes underperforming its best constituent path. Our work provides systematic, quantitative evidence of this ``weakest link'' phenomenon in complex hybrid search, challenging the common ``more is better'' assumption.

The primary implication is that hybrid architectures are sensitive systems, not simple ensembles where strengths automatically aggregate. This highlights the symbiosis between candidate generation and re-ranking: a re-ranker's effectiveness is fundamentally constrained by its inputs. It cannot recover documents missed during initial retrieval, making candidate quality a hard ceiling on final accuracy. A critical design principle is therefore to validate component performance before integration. Dynamic path selection is a potential mitigation, which we will revisit in Section \ref{sec: future_work}.

\subsubsection{A Data-Driven Map of Performance Trade-offs}
Table \ref{tab:qualitative_summary} shows that architectures excel in specific dimensions (e.g., {\ts}-inclusive paths for accuracy, single {\dvs} for speed). This reveals a fundamental trade-off: no single architecture excels across accuracy, efficiency, and resource overhead simultaneously. Currently, navigating these trade-offs has been challenging, with choices often based on intuition or ad-hoc pairwise studies. Our work provides a more systematic foundation, enabling quantitative assessment of a configuration's specific performance profile.

The practical implication is that a static architecture is likely suboptimal. Our findings instead point to the need for systems that adaptively select path combinations based on query type, data, or performance requirements. Our trade-off map also reveals effective patterns, such as the {\fts}+{\svs}+{\dvs} combination, which emerges as a highly balanced ``sweet spot'': it achieves high accuracy without the severe resource demands of {\ts}-inclusive configurations.

\subsubsection{TRF: A Practical Approach for In-Database Fine-Grained Re-ranking}
We find that Tensor-based Re-ranking Fusion (TRF) is a practical and highly effective architecture for in-database re-ranking. It fills a key gap in the re-ranking landscape: our results (Section \ref{subsec: exp rq3}) show it consistently outperforms coarse-grained fusion methods like RRF and WS. Unlike a full {\ts} path, TRF operates only on a small candidate set, making it far more resource-efficient. Unlike heavyweight cross-encoders \cite{ZhangZLXDTLYXHZ24,KhattabZ20} requiring separate GPU processing, TRF is designed as an in-database operation.

The implication is that TRF offers a clear upgrade path from traditional fusion methods. Practitioners can gain a significant accuracy boost for a moderate increase in latency and memory, making TRF attractive for accuracy-critical applications. We highlight that these strong results were achieved with a baseline, unoptimized TRF implementation; its high performance, coupled with optimization potential (e.g., tensor compression), marks it as a promising direction for future database retrieval systems, as we discuss next.

\subsection{Challenges and Future Directions}
\label{sec: future_work}
Our findings illuminate key challenges and open promising research directions. We articulate five: developing adaptive architectures, managing index maintenance costs, optimizing tensor-based re-ranking, extending hybrid search to multi-modal documents, and performing end-to-end evaluations within RAG systems.

\subsubsection{Towards Adaptive Hybrid Search}
Our findings---that a ``weakest link'' can degrade performance and that no single architecture is universally optimal in performance trade-offs---reveal a critical limitation of current static systems. A fixed combination of retrieval paths is unlikely to be optimal for all queries or data subsets. This highlights a major research direction: \textit{adaptive hybrid search architectures}. Future systems could incorporate a cost-based query optimizer to predict the effectiveness and resource cost of each available path, enabling the dynamic selection of an optimal path set based on user-defined constraints (e.g., a latency budget) and pruning a predicted ``weak link'' at runtime. Moving from static to query-aware execution is a crucial step for building robust hybrid search systems.
For example, our analysis (Section \ref{sec:effectiveness_results}) indicates query length is a simple yet effective heuristic: short, keyword-centric queries are often best served by \fts{}'s low-latency precision, whereas longer, descriptive queries provide the context for \dvs{} to achieve superior semantic matching. A query-aware routing mechanism analyzing these characteristics could dynamically select an optimal path combination, mitigating the ``weakest link'' problem.

\subsubsection{Managing Index Maintenance Costs}
A further evaluation may consider not only read-path performance but also the cost of maintaining multiple indexes. Our evaluation (Figure \ref{fig: index overhead}) quantifies the initial, offline construction costs, detailing import time, construction time, peak memory, and on-disk size for each paradigm. This provides a view of static deployment costs.
However, real-world data is rarely static. A critical area is the analysis of \textit{dynamic} maintenance costs, such as ingestion throughput and update/delete latency. Architectural choices benefiting read performance may introduce significant write-path complexities; for instance, maintaining consistency across four separate indexes during high-volume updates is a non-trivial challenge. A systematic study of these dynamic trade-offs is essential for understanding the total cost of ownership of advanced hybrid search systems.

\subsubsection{Optimizing the Cost-Performance of Tensor-based Re-ranking}
Our finding identifies TRF as a high-efficacy method that outperforms traditional fusion. Its broader adoption, however, is limited by higher computational and memory costs compared to RRF. This presents a valuable research direction: \textit{optimizing the cost} of TRF. Tensor compression is a promising avenue; as our follow-up work suggests, techniques like scalar or binary quantization can dramatically reduce the memory footprint of tensor representations while largely preserving re-ranking effectiveness. Complementing this, computation acceleration for the MaxSim operation is critical, perhaps by using hardware-specific instructions (e.g., SIMD) or developing more efficient token-level interaction algorithms. Such optimizations could eliminate the primary barriers to TRF's adoption, positioning it as a new standard for high-accuracy re-ranking.

\subsubsection{Hybrid Search for Multi-Modal Documents}
While our work provides a foundation for textual hybrid search, a significant future direction is extending these principles to \textit{multi-modal documents}. This requires designing fusion mechanisms that can effectively weigh and combine signals from visual modalities (e.g., semantic embeddings from models like ColPali~\cite{FaysseSWOVHC25}) with those from textual modalities (e.g., exact keyword matches from Optical Character Recognition (OCR)~\cite{Smith07}) to improve retrieval performance.

\subsubsection{End-to-End Evaluation within RAG Systems}
Another vital direction is evaluating hybrid search \textit{within an end-to-end Retrieval-Augmented Generation (RAG) framework}, not just as standalone components. The retrieval module's performance is fundamentally coupled with other pipeline components. A system-level evaluation must therefore analyze the interplay between the chosen hybrid strategy and other factors, such as document chunking policies and query intent recognition. Understanding how these architectural interactions impact the quality and factuality of the final generated output is essential for building optimized RAG systems.

%% file: sections/7_conclusion.tex
\section{Conclusion}
\label{sec: conclusion}
We conducted the first systematic study of advanced hybrid search architectures. Our novel evaluation framework comprehensively evaluates four major retrieval paradigms and their 15 combinations across 11 real-world datasets. Our analysis yielded three principal findings: (1) the ``weakest link'' phenomenon, where a system's accuracy is constrained by its least effective component; (2) a data-driven map of the performance landscape, showing that architectural selection is a multi-dimensional trade-off with no universal solution; and (3) the emergence of TRF as a promising re-ranking architecture. We believe our results provide a crucial foundation for practitioners building next-generation retrieval systems and for researchers exploring adaptive hybrid search.